\documentclass[aps,pra,twocolumn,showpacs]{revtex4}
\usepackage{graphicx}
\usepackage{psfrag}
\usepackage{amsmath}
\usepackage{amssymb}
\usepackage{bm}

\renewcommand{\vec}[1]{\boldsymbol{#1}}
\renewcommand{\tensor}[1]{{\boldsymbol{\mathit {#1}}}}
\newcommand{\Nabla}{\vec{\nabla}}
\newcommand{\Lnabla}{\overset{\leftarrow}{\Nabla}}

\newcommand{\EW}[1]{\langle #1 \rangle}
\newcommand{\EWb}[1]{\bigl\langle #1 \bigr\rangle}
\newcommand{\D}{\mathrm{d}}

\renewcommand{\Im}{{\rm Im}\,}
\renewcommand{\Re}{{\rm Re}\,}

\newcommand{\fo}[1]{\underline{\hat{\mathbf{#1}}}}

\begin{document}
\title{Casimir force acting on magnetodielectric bodies embedded in media}
\author{Christian Raabe}
\author{Dirk-Gunnar Welsch}
\affiliation{Theoretisch-Physikalisches Institut,
Friedrich-Schiller-Universit\"at Jena, Max-Wien-Platz 1, D-07743 Jena,
Germany}
\date{January 4, 2005}
\begin{abstract}
Within the framework
of macroscopic quantum electrodynamics, general expressions for the
Casimir force acting on linearly and causally responding
magnetodielectric bodies that can be embedded in another linear
and causal magnetodielectric medium are derived.
Consistency with microscopic harmonic-oscillator
models of the matter is shown.
The theory is applied to planar structures and
proper generalizations of Casimir's and Lifshitz-type
formulas are given.
\end{abstract}
\maketitle
\section{Introduction}
\label{sec1}
It is well known that the introduction
of momentum and energy of the macroscopic electromagnetic field
requires careful consideration, even for linear media.
In fact, insertion of constitutive relations into Gauss's
and Ampere's laws prevents one, in general, 
from deriving local balance equations
of a similar type as in the microscopic theory.
However, with quite restrictive
(and most questionable) assumptions about the
material under consideration,
these difficulties can be formally overcome.
Therefore, textbooks typically resort to approximate formulas
that are based on assumptions such as
quasimonochromatic fields
and lossless media \cite{LanLifED,SchwingerED}.
Although the limitations inherent in such theories
are rather obvious, they are nevertheless applied
beyond their range of validity.

A typical example is the Casimir effect, which is closely
related to the changes in the vacuum electromagnetic-field energy
and/or momentum flow (stress) induced by the presence of
inhomogeneous matter. With regard to the calculation of
the Casimir force acting on macroscopic bodies that are embedded 
in a medium, the question of what are the correct expressions
for these quantities becomes crucial.
Frequently, expressions that seem reasonable
at first glance---such as Minkowski's stress tensor---have
been taken for granted without justification.
As we shall see, this has led to incorrect
extensions of the well-known Lifshitz formula for
the Casimir force
between two dielectric half-spaces separated by vacuum
to the case where the interspace is not empty but
also filled with material
\cite{Schwinger,TomasCasimir,ZhouF071995} (see also the textbooks
\cite{AbrikosovStatPhys,GinzburgED,MilonniQuantumVacuum}
and references therein).

In this paper, we reconsider, within the framework
of macroscopic quantum electrodynamics, the problem of the
calculation of Casimir forces, by regarding
the Lorentz force density as the fundamental quantity.
The Lorentz force acting on some (macroscopic)
spatial region containing (electrically neutral)
matter is of course the corresponding volume integral of the
Lorentz force density, where the relevant charge and current
densities may be thought of as being
expressed in terms of the polarization 
and the magnetization of the matter.
As a consequence, the Casimir force on a macroscopic body and, 
equivalently, the stress on its surface
can be expressed---in close analogy with microscopic
electrodynamics---in terms of the electric and induction fields,
irrespective of any specific constitutive relations.
In particular, if the body linearly responds to the electric and 
induction fields and the medium the body is embedded
in is also a linear one, then the Casimir force can be expressed
solely in terms of the classical (retarded) Green tensor,
which in turn is determined by the response functions of 
the magnetodielectric matter under consideration.

We show that the Casimir force formula found in this way is 
consistent with microscopic theories based on harmonic-oscillator
models of dispersing and absorbing dielectric matter. The formula is
valid under very general conditions, and enables one to 
study in a consistent way the Casimir force on linearly responding,
dispersing, and absorbing magnetodielectric bodies that are not 
necessarily placed in vacuum but may also be surrounded by a 
dispersing and absorbing linear magnetodielectric medium.
Since magnetodielectric matter is included in the theory, 
it is possible to consider also left-handed materials
\cite{VeselagoVG071964}. To illustrate the theory, we apply it
to a planar geometry and derive a proper extension of
Lifshitz-type formulas, with emphasis also on
the extension of Casimir's original formula.

The paper is organized as follows. In Sec.~\ref{sec2}, the
stress tensor associated with the
(macroscopic) Lorentz force is introduced.
The Casimir force
is calculated in Sec.~\ref{sec3}, and
Sec.~\ref{sec4} makes contact with the microscopic 
harmonic-oscillator model. The application of the theory 
to planar structures is given in Sec.~\ref{sec5}, and a
summary and some concluding remarks are given in Sec.~\ref{sec6}.
       
\section{Lorentz force and stress tensor}
\label{sec2}

Let us begin with the classical Maxwell equations
for the electric and induction fields
$\mathbf{E}$ and $\mathbf{B}$ in the presence of matter,
\begin{align}
\label{Max1}
\Nabla \mathbf{B}&=0,
\\
\label{Max2}
\Nabla\times\mathbf{E} + \frac{\partial\mathbf{B}}{\partial t} &=0,
\end{align}
\begin{align}
\label{Max3}
\varepsilon_0 \Nabla \mathbf{E} &=\rho,\\
\label{Max4}
\mu_{0}^{-1}\Nabla\times\mathbf{B} -
\varepsilon_0\frac{\partial\mathbf{E}}{\partial t} &= \mathbf{j}.
\end{align}
In this paper, ``dot products'' are written without the dot, and dyadic
products are denoted by $\otimes$.
In Eqs.~(\ref{Max3}) and (\ref{Max4}), $\rho$ and $\mathbf{j}$
cover \emph{all} charges and currents of the
system
under consideration.
Within the framework of a macroscopic description,
the ``internal'' charges and currents associated with the
particles that form some neutral
material system are commonly described in terms of polarization 
and magnetization fields $\mathbf{P}$ and $\mathbf{M}$, respectively. 
The remaining ``external'' charges and currents---if any---are
kept explicitly, i.e.,
\begin{gather}
\label{eq5}
\rho = \rho_\mathrm{int} + \rho_\mathrm{ext},\\
\label{eq6}
\mathbf{j} = \mathbf{j}_\mathrm{int} + \mathbf{j}_\mathrm{ext},
\end{gather} 
where
\begin{gather}
\label{Max5}
\rho_\mathrm{int}=
- \Nabla \mathbf{P},\\
\label{Max6}
\mathbf{j}_\mathrm{int} =
\frac{\partial\mathbf{P}}{\partial t} + \Nabla\times \mathbf{M}.
\end{gather} 
As long as constitutive equations (relating
the polarization and magnetization fields to the electric and 
induction fields) are not introduced, 
Eqs.~(\ref{Max1})--(\ref{Max6}),
which may also be interpreted microscopically,
are generally valid. Clearly,
the defining equations
(\ref{Max5}) and (\ref{Max6}) of $\mathbf{P}$ and $\mathbf{M}$,
respectively, can be satisfied
for any choice of (conserved) ``internal'' sources,
the corresponding integrability condition being just
\begin{equation}
\label{eq9}
\frac{\partial\rho_\mathrm{int}}{\partial t} +
\Nabla\mathbf{j}_\mathrm{int}=0.
\end{equation}
Note that the ``internal'' sources typically comprise
bound charges and the associated currents---a concept that
is commonly used together with spatial averaging
in macroscopic electrodynamics \cite{Jackson3rd}.

As known, the Lorentz force density
\begin{equation}
\label{Max7}
\mathbf{f}=\rho \mathbf{E} + \mathbf{j}\times \mathbf{B}
\end{equation}
can be rewritten
with the help of Eqs.~(\ref{Max1})--(\ref{Max4}) as
\begin{equation}
\label{Max9}
\mathbf{f}=\Nabla \tensor{T}
-\varepsilon_{0}\frac{\partial}{\partial t}
(\mathbf{E} \times \mathbf{B}),
\end{equation}
where the stress tensor
\begin{equation}
\label{Max8}
\tensor{T}= \varepsilon_0 \mathbf{E \otimes E} + \mu_0^{-1}
\mathbf{B \otimes B} - \textstyle{\frac{1}{2}} (\varepsilon_0 E^2 +\mu_0^{-1}
B^2)\tensor{1}
\end{equation}
has been introduced ($\tensor{1}$, unit tensor).
Clearly, Eqs.~(\ref{Max7}) and (\ref{Max9}) are universally valid,
regardless of whether the charge and current densities
have been decomposed according to Eqs.~(\ref{eq5})--(\ref{Max6})
or not. Note that essentially the same position
has been recently taken up \cite{Obukhov} in the (re)analysis
of measurements of the electromagnetic force that acts on dielectric
\cite{WalkerAndWalkerExperiment1,WalkerAndWalkerExperiment2,%
WalkerAndWalkerExperiment3,WalkerAndWalkerExperiment4,WalkerAndWalkerExperiment5}
or magnetodielectric
\cite{JamesExperiment} (see also Refs.~\cite{JamesExperiment2,BrevikReport})
disks exposed to crossed electric and magnetic fields.
(For a different perspective, see also Ref.~\cite{GarrisonJC062004}).

The integral of the Lorentz force density $\mathbf{f}$ over some 
space region (volume $V$) gives of course the total 
electromagnetic force $\mathbf{F}$ acting on the matter inside it,
\begin{equation}
\label{eq12}
\mathbf{F} = \int_{V} \D^3r\,\mathbf{f}.
\end{equation}
Using Eq.~(\ref{Max9}), we have
\begin{equation}
\label{eq13}
\mathbf{F}
= \int_{\partial V}\D\mathbf{a} \tensor{T}
-\varepsilon_0\frac{\D}{\D t} \int_V \D^3r\,\mathbf{E}\times\mathbf{B},
\end{equation}
which is obviously also true if the space region is
occupied by a macroscopic body, with the charges and currents
being ``internal'' ones described by polarization and magnetization fields.
In particular, if the volume integral on the right-hand side
of this equation does not depend on time, then the total
force reduces to the surface integral
\begin{equation}
\label{eq14}
\mathbf{F}
= \int_{\partial V} \D\mathbf{F},
\end{equation}
where
\begin{equation}
\label{eq15}
\D\mathbf{F} = \D\mathbf{a} \tensor{T}
= \tensor{T} \D\mathbf{a}
\end{equation}
may be regarded as the infinitesimal force element
acting on an infinitesimal surface element $\D\mathbf{a}$. 
Note that a constant term in the stress tensor does not contribute 
to the integral in Eq.~(\ref{eq14}) and can therefore be omitted.
In the calculation of the Casimir force in Sec.~\ref{sec3},
it will be necessary to make use of this fact.
 
Expressing in Eq.~(\ref{eq15}) the stress tensor
$\tensor{T}$ in terms of
Minkowski's stress tensor $\tensor{T}^\mathrm{(M)}$
(which agrees with Abraham's stress tensor \cite{GinzburgED}),
%
\begin{eqnarray}
\label{Max11}
\lefteqn{
\tensor{T}^\mathrm{(M)}
= \mathbf{D \otimes E} + \mathbf{H \otimes B}
- \textstyle{\frac{1}{2}} (\mathbf{D E} + \mathbf{H B})\tensor{1}
}
\nonumber\\&&
= \tensor{T}
+ \mathbf{P} \otimes \mathbf{E} -
\mathbf{M} \otimes \mathbf{B}
- \textstyle{\frac{1}{2}} (\mathbf{P}\mathbf{E} -
\mathbf{M}\mathbf{B})
\tensor{1},
\quad
\end{eqnarray}
one finds that 
\begin{equation}
\label{Max13}
\D\mathbf{F}
= \D\mathbf{a}\tensor{T}^\mathrm{(M)} -
\D\mathbf{a}\left[\mathbf{P} \otimes \mathbf{E} -
\mathbf{M} \otimes \mathbf{B}
- \textstyle{\frac{1}{2}} (\mathbf{P}\mathbf{E} -
\mathbf{M}\mathbf{B})\right],
\end{equation}
from which it is seen that in general
\begin{equation}
\label{eq18}
\D\mathbf{F} \neq \D\mathbf{a} \tensor{T}^\mathrm{(M)}.
\end{equation}
That is to say, the use of Minkowski's
stress tensor is expected not to yield the correct force in general,
whereas the use of $\tensor{T}$, which is formally
the same as the stress tensor in microscopic electrodynamics,
is \emph{always} correct.

Let
\begin{gather}
\label{Max14}
\mathbf{P} =
\mathbf{P}_\mathrm{ind}
+\mathbf{P}_\mathrm{N},\\
\label{eq20}
\mathbf{M} =
\mathbf{M}_\mathrm{ind}
+\mathbf{M}_\mathrm{N}
\end{gather}
be the decompositions of the polarization and the magnetization
into induced parts $\mathbf{P}_\mathrm{ind}$, $\!\mathbf{M}_\mathrm{ind}$
and noise parts $\mathbf{P}_\mathrm{N}$,$\mathbf{M}_\mathrm{N}$,
where the noise parts are closely related to dissipation.
Substituting in Eq.~(\ref{Max13}) for $\mathbf{P}$ and
$\mathbf{M}$ the expressions (\ref{Max14}) and (\ref{eq20}),
respectively, we see that force calculations
that are based on Minkowski's stress tensor are 
expected to be 
incorrect with respect to both the induced parts and the
noise parts of the polarization and the magnetization in general.
Clearly, if---and only if---the aim is to calculate
the force acting on bodies that are placed in a free-space region,
then both $\tensor{T}$ and $\tensor{T}^\mathrm{(M)}$
lead to the same result.

The idea to regard [according to Eqs.~(\ref{Max7})--(\ref{eq12})]
the Lorentz force acting on the totality of charges and
currents belonging to a system under consideration
as the fundamental quantity
is neither new \cite{Obukhov,Nelson,Poincelot,Livens}
nor particularly hard to agree with.
Despite this, the use of Minkowski's stress tensor or related quantities has
still been common in the calculation of electromagnetic forces.
In this context, let us make a few general remarks.
The momentum that may be introduced on the basis of
Eq.~(\ref{Max9}) is related to the Noether symmetry expressing
homogeneity of space. It must be distinguished from the 
pseudomomentum related to (strict) homogeneity of the material.
In connection with the so-called Minkowski-Abraham controversy,
Refs.~\cite{Nelson,LoudonAllenNelson} analyze in a Lagrangian
framework the meaning of different momentumlike quantities
by consideration of explicit (classical) dynamical models of a 
homogeneous dielectric. In Ref.~\cite{Nelson}, the
homogeneous dielectric is assumed to be lossless and treated in some
multipolar, long-wavelength approximation (for an inclusion
of magnetic properties, see \cite{NelsonChen}).
In Ref.~\cite{LoudonAllenNelson}, the homogeneous dielectric
is described by a single-resonance Drude-Lorentz model.
All the calculations show that Eq.~(\ref{Max9})
[together with Eqs.~(\ref{Max7}) and (\ref{Max8})] is really
the momentum balance of the macroscopic electromagnetic field.

\section{Casimir force on bodies embedded in
dispersing and absorbing media}
\label{sec3}

In classical electrodynamics, electrically neutral
material bodies at zero temperature which do not carry a
permanent polarization and/or magnetization 
are not subject to a Lorentz force in the absence of
external electromagnetic fields. As known, the situation
changes in quantum electrodynamics, since the
vacuum fluctuations of the electromagnetic field can
give rise to a nonvanishing Lorentz force---the Casimir force. 
Its experimental demonstration has therefore
been regarded as a confirmation of quantum theory.

To translate the classical formulas given in Sec.~\ref{sec2} into 
the language of quantum theory, let us consider linear, inhomogeneous 
media that locally respond to the electromagnetic field and
can thus be characterized by a spatially varying
complex permittivity $\varepsilon(\mathbf{r},\omega)$
and a spatially varying complex permeability $\mu(\mathbf{r},\omega)$.
Following Ref.~\cite{Welsch}, we may write the medium-assisted electric
and induction field operators in the form of    
\begin{gather}
\label{2.4b}
\hat{\mathbf{E}}(\mathbf{r})
= \int_0^\infty \D\omega\, \fo{E}(\mathbf{r},\omega) + \mathrm{H.c.},
\\
\label{2.4b-1}
\hat{\mathbf{B}}(\mathbf{r})
= \int_0^\infty \D\omega\, \fo{B}(\mathbf{r},\omega) + \mathrm{H.c.},
\end{gather}
where
\begin{gather}
\label{2.2}
\fo{E}(\mathbf{r},\omega)=i\mu_{0}\omega\int
\D^3r'\,\tensor{G}(\mathbf{r,r'},\omega)
\fo{j}_{\mathrm N}(\mathbf{r}',\omega),
\\
\label{2.2a}
\fo{B}(\mathbf{r,\omega})=\mu_{0}
\Nabla\times \int \D^3r'\,\tensor{G}(\mathbf{r,r'},\omega)
\fo{j}_{\mathrm N}(\mathbf{r}',\omega).
\end{gather}
Here, $\tensor{G}(\mathbf{r,r'},\omega)$ is the classical Green tensor, 
which has to be determined from the equation
\begin{equation}
\label{2.1}
\Nabla\times\kappa(\mathbf{r},\omega)\!\Nabla\times
\tensor{G}(\mathbf{r,r'},\omega)
-
\frac{\omega^2}{c^2}\varepsilon(\mathbf{r},\omega)
\!\tensor{G}(\mathbf{r,r'},\omega)
\!=\!\tensor{\delta}(\mathbf{r\!-\!r'})
\end{equation}
together with the boundary condition at infinity, and
$\fo{j}_{\mathrm N}(\mathbf{r},\omega)$ is defined by
\begin{equation}
\label{2.4}
\fo{j}_\mathrm{N}(\mathbf{r},\omega) =
-i\omega \fo{P}_\mathrm{N}(\mathbf{r},\omega)
+\Nabla\times\fo{M}_\mathrm{N}(\mathbf{r},\omega),
\end{equation}
where $\fo{P}_\mathrm{N}(\mathbf{r},\omega)$ and
$\fo{M}_\mathrm{N}(\mathbf{r},\omega)$ are, respectively,
the (fluctuating) noise parts of the polarization
$\fo{P}(\mathbf{r},\omega)$  and the magnetization 
$\fo{M}(\mathbf{r},\omega)$ in the frequency domain, 
\begin{eqnarray}
\label{2.3}
\fo{P}(\mathbf{r},\omega)&=&\varepsilon_{0}
[\varepsilon(\mathbf{r},\omega)-1]\fo{E}(\mathbf{r},\omega)
+\fo{P}_\mathrm{N}(\mathbf{r},\omega),
\\
\label{2.3-1}
\fo{M}(\mathbf{r},\omega)&=&\kappa_{0}[1-\kappa(\mathbf{r},\omega)]\fo{B}
(\mathbf{r},\omega)+\fo{M}_\mathrm{N}(\mathbf{r},\omega),
\end{eqnarray}
[$\kappa_{0}$ $\!=$ $\!\mu_{0}^{-1}$,
$\kappa(\mathbf{r},\omega)$ $\!=$ $\!\mu^{-1}(\mathbf{r},\omega)$].

The Green tensor [as well as $\varepsilon(\mathbf{r},\omega)$ and 
$\kappa(\mathbf{r},\omega)$] is holomorphic in the upper
$\omega$ half-plane and has the ``reality'' property
\begin{equation}
\label{2.4d}
\tensor{G}(\mathbf{r,r'},-\omega^{\ast})
=\tensor{G}^{\ast}(\mathbf{r,r'},\omega).
\end{equation}
Moreover, it obeys the reciprocity relation 
\begin{equation}
\label{2.4c}
\tensor{G}(\mathbf{r,r'},\omega)
=\tensor{G}^\mathsf{T}(\mathbf{r',r},\omega)
\end{equation}
(the superscript $\mathsf{T}$ denotes matrix transposition) and the
integral relation
\begin{eqnarray}
\label{2.5}
\lefteqn{
\int \D^3s\, \Bigl\{      
\bigl[\tensor{G}(\mathbf{r,s},\omega)\!\times\! \Lnabla_{\!s}\bigr]
\Im\kappa(\mathbf{s},\omega)
\bigl[\Nabla_{\!s}\!\times\!\tensor{G}^{\ast}(\mathbf{s,r'},\omega)\bigr]
}
\nonumber\\&&\hspace{-1ex}
+\,\frac{\omega^2}{c^2}\tensor{G}(\mathbf{r,s},\omega)
\Im\varepsilon(\mathbf{s,\omega})
\tensor{G}^{\ast}(\mathbf{s,r'},\omega)
\Bigr\}\!=\!\Im\tensor{G}(\mathbf{r,r'},\omega),
\nonumber\\&&
\end{eqnarray}
where the notation
\begin{equation}
\label{2.6}
\tensor{G}(\mathbf{r,r'},\omega)\times\Lnabla{'}
=-[\Nabla'\times \tensor{G}^\mathsf{T}
(\mathbf{r,r'},\omega)]^{\mathsf T}
\end{equation}
has been used.

According to Ref.~\cite{Welsch},
$\fo{P}_\mathrm{N}(\mathbf{r},\omega)$ and
$\fo{M}_\mathrm{N}(\mathbf{r},\omega)$
can be related to bosonic fields $\hat{\mathbf{f}}_{e}(\mathbf{r},\omega)$
and $\hat{\mathbf{f}}_{m}(\mathbf{r},\omega)$,
respectively, in such a way that the correct (equal-time)
commutation relations of the electromagnetic field operators
are satisfied,
\begin{gather}
\label{2.4-1}
\fo{P}_\mathrm{N}(\mathbf{r},\omega)
=i\left[\hbar\varepsilon_{0}
\Im\varepsilon(\mathbf{r,\omega})/\pi\right]^{1/2}
\hat{\mathbf{f}}_{e}(\mathbf{r},\omega),
\\
\label{2.4-2}
\fo{M}_\mathrm{N}(\mathbf{r},\omega)
=\left[-\hbar\kappa_{0}
\Im\kappa(\mathbf{r,\omega})/\pi\right]^{1/2}
\hat{\mathbf{f}}_{m}(\mathbf{r},\omega),
\end{gather}
\begin{equation}
\label{2.4-3}
\bigl[\hat{f}_{\lambda k}(\mathbf{r},\omega),
\hat{f}_{\lambda'l}^\dagger(\mathbf{r},\omega)\bigr]
= \delta_{kl}\delta_{\lambda\lambda'}\delta(\mathbf{r}-\mathbf{r}')
\delta(\omega-\omega')
\end{equation}
($\lambda$ $\!=$ $\!e,m$).
Note that the $\hat{\mathbf{f}}_{\lambda}(\mathbf{r},\omega)$
and $\hat{\mathbf{f}}_{\lambda}^\dagger(\mathbf{r},\omega)$
play the role of the dynamical
(canonical) variables of the theory.
Combining Eqs.~(\ref{2.4b})--(\ref{2.2a}) with
Eqs.~(\ref{2.4}), (\ref{2.4-1}), and (\ref{2.4-2})
yields the electric and induction fields in terms
of the dynamical variables.  

The charge and current densities that are subject to
the Lorentz force are given by
\begin{gather}
\label{eq34}
\hat{\rho}(\mathbf{r})
= \int_0^\infty \D\omega\,\hat{\underline{\rho}}(\mathbf{r},\omega)
+ \mathrm{H.c.},
\\
\label{eq35}
\hat{\mathbf{j}}(\mathbf{r})
= \int_0^\infty \D\omega\,\hat{\underline{\mathbf{j}}}(\mathbf{r},\omega)
+ \mathrm{H.c.},
\end{gather}
where, according to Eqs.~(\ref{eq5})--(\ref{Max6})
($\hat{\rho}_\mathrm{ext}$ $\!=$ $\!0$,
$\hat{\mathbf{j}}_\mathrm{ext}$ $\!=$ $\!0$)
together with Eqs.~(\ref{2.4}), (\ref{2.3}), and (\ref{2.3-1}),
\begin{equation}
\label{eq36}
\fo{\rho}(\mathbf{r},\omega) =
-\varepsilon_{0}
\Nabla\{[\varepsilon(\mathbf{r},\omega)-1]\fo{E}(\mathbf{r},\omega)\}
+ (i\omega)^{-1} \Nabla\fo{j}_\mathrm{N}(\mathbf{r},\omega)
\end{equation}
and
\begin{eqnarray}
\label{eq37}
\lefteqn{
\fo{j}(\mathbf{r},\omega)
=-i\omega \varepsilon_{0}
[\varepsilon(\mathbf{r},\omega)-1]
\fo{E}(\mathbf{r},\omega)
}
\nonumber\\&&
+\,\Nabla\times \{\kappa_{0}[1-\kappa(\mathbf{r},\omega)]
\fo{B}(\mathbf{r},\omega)\}
+\fo{j}_\mathrm{N}(\mathbf{r},\omega).
\quad
\end{eqnarray}
Using the definitions of $\fo{E}(\mathbf{r},\omega)$,
$\fo{B}(\mathbf{r},\omega)$, and
$\fo{j}_\mathrm{N}(\mathbf{r},\omega)$ together with the
bosonic commutation relations for the fundamental fields
$\hat{\mathbf{f}}_\lambda(\mathbf{r},\omega)$ and
$\hat{\mathbf{f}}^\dagger_\lambda(\mathbf{r},\omega)$,
one can prove (Appendix~\ref{appC}) that
\begin{align}
\label{eq37-1}
\bigl[\hat{\rho}(\mathbf{r}),
\hat{\mathbf{E}}(\mathbf{r}')\bigr]
=0
,\\
\label{eq37-2}
\bigl[\hat{\rho}(\mathbf{r}),
\hat{\mathbf{B}}(\mathbf{r}')\bigr]
= 0,\\
\label{eq37-3}
\bigl[\,\hat{\!j}_k(\mathbf{r}),
\hat{B}_{l}(\mathbf{r}')\bigr]
= 0,
\end{align}
and
\begin{equation}
\label{eq37-4}
\bigl[\,\hat{\!j}_k(\mathbf{r}),
\hat{E}^\perp_{l}(\mathbf{r}')\bigr]
=i\hbar\,\Omega_{\varepsilon}^2(\mathbf{r})
{\delta}^{\perp}_{kl}(\mathbf{r-r'}),
\end{equation}
where the position-dependent plasma frequency
$\Omega_{\varepsilon}(\mathbf{r})$
is defined by the asymptotic behavior of the permittivity
for large $\omega$ in the upper half-plane
according to
$\varepsilon(\mathbf{r,\omega})
\!\simeq\!1\!-\!\Omega_{\varepsilon}^2(\mathbf{r})/\omega^2$.
The commutation relations (\ref{eq37-1})--(\ref{eq37-4})
clearly show that $\hat{\rho}(\mathbf{r})$
and $\hat{\mathbf{j}}(\mathbf{r})$ really represent
matter quantities. It is worth noting that
Eq.~(\ref{eq37-4}) exactly corresponds
to the equation obtained when---on the basis of a microscopic
description---the current density is explicitly specified 
in terms of particle velocities (Appendix~\ref{appC}).

If the field-matter system is in a number state [defined
with respect to the number (density) operators
$\hat{\mathbf{f}}_{\lambda}^{\dagger}(\mathbf{r},\omega)
\hat{\mathbf{f}}_{\lambda}(\mathbf{r},\omega)$]
such as the ground state, or an incoherent mixture of them
such as a thermal state, then all one-time averages are 
evidently time-independent. Recalling the bosonic character 
of the fundamental fields $\hat{\mathbf{f}}_{\lambda}(\mathbf{r,\omega})$
[and $\hat{\mathbf{f}}_{\lambda}^\dagger(\mathbf{r,\omega})$]
and assuming them to be excited in thermal states, we easily 
obtain, in close analogy to Ref.~\cite{Raabe1}, 
\begin{multline}
\label{eq39}
\bigl\langle\hat{\mathbf{f}}_{\lambda}(\mathbf{r,\omega})\otimes
\hat{\mathbf{f}}_{\lambda'}^{\dagger}
(\mathbf{r',\omega'})\bigr\rangle
\\
={\textstyle\frac{1}{2}}
\left[\coth\!\left(\frac{\hbar\omega}{2k_\mathrm{B}T}\right)+1
\right]\delta_{\lambda\lambda'}\delta(\omega-\omega')
\tensor{\delta}(\mathbf{r-r'}),
\end{multline}
\begin{multline}
\label{eq40}
\bigl\langle\hat{\mathbf{f}}_{\lambda}^{\dagger}(\mathbf{r,\omega})\otimes
\hat{\mathbf{f}}_{\lambda'}
(\mathbf{r',\omega'})\bigr\rangle
\\
={\textstyle\frac{1}{2}}
\left[\coth\!\left(\frac{\hbar\omega}{2k_\mathrm{B}T}\right)-1
\right]\delta_{\lambda\lambda'}\delta(\omega-\omega')
\tensor{\delta}(\mathbf{r-r'}),
\end{multline}
\begin{equation}
\label{eq41}
\bigl\langle\hat{\mathbf{f}}_{\lambda}(\mathbf{r,\omega})\otimes
\hat{\mathbf{f}}_{\lambda'}
(\mathbf{r',\omega'})\bigr\rangle = 0.
\end{equation}
Making use of Eq.~(\ref{2.4}) together with Eqs.~(\ref{2.4-1})
and (\ref{2.4-2}), we find that the correlation functions
(\ref{eq39})--(\ref{eq41}) imply the correlation functions
\begin{eqnarray}
\label{2.8}
\lefteqn{
\bigl\langle\fo{\mathbf{j}}_\mathrm{N}(\mathbf{r,\omega})\otimes
\fo{\mathbf{j}}_\mathrm{N}^{\dagger}
(\mathbf{r',\omega'})\bigr\rangle
}
\nonumber\\&&\hspace*{-1ex}
=\frac{\hbar}{2\pi\mu_{0}}\,\delta(\omega-\omega')
\left[\coth\!\left(\frac{\hbar\omega}{2k_\mathrm{B}T}\right)\!+\!1
\right]
\nonumber\\&&\hspace*{-1ex}
\times \,
\left\{\frac{\omega^2}{c^2}
\sqrt{\Im \varepsilon(\mathbf{r},\omega)}
\tensor{\delta}(\mathbf{r-r'})
\sqrt{\Im\varepsilon(\mathbf{r'},\omega')}
\right.\nonumber\\&&\hspace*{-1ex}\left.
+\,\Nabla\!\times\!
\left[
\sqrt{\Im \kappa(\mathbf{r},\omega)}
\tensor{\delta}(\mathbf{r-r'})
\sqrt{\Im\kappa(\mathbf{r'},\omega')}
\right]
\!\times\!\Lnabla{'}
\right\}\!,
\qquad
\end{eqnarray}
\begin{eqnarray}
\label{2.8a}
\lefteqn{
\bigl\langle\fo{\mathbf{j}}_\mathrm{N}^{\dagger}(\mathbf{r,\omega})\otimes
\fo{\mathbf{j}}_\mathrm{N}
(\mathbf{r',\omega'})\bigr\rangle
}\nonumber\\&&\hspace*{-1ex}
=\frac{\hbar}{2\pi\mu_{0}}\,\delta(\omega-\omega')
\left[\coth\!\left(\frac{\hbar\omega}{2k_\mathrm{B}T}\right)\!-\!1
\right]
\nonumber\\&&\hspace*{-1ex}
\times \,
\left\{\frac{\omega^2}{c^2}
\sqrt{\Im \varepsilon(\mathbf{r},\omega)}
\tensor{\delta}(\mathbf{r-r'})
\sqrt{\Im
\varepsilon(\mathbf{r'},\omega')}
\right.\nonumber\\&&\hspace*{-1ex}\left.
+\,\Nabla\!\times\!
\left[
\sqrt{\Im \kappa(\mathbf{r},\omega)}
\tensor{\delta}(\mathbf{r-r'})
\sqrt{\Im
\kappa(\mathbf{r'},\omega')}
\right]
\!\times\!\Lnabla{'}
\right\}\!,
\qquad
\end{eqnarray}
and
\begin{equation}
\label{2.8b}
\EWb{\fo{\mathbf{j}}_\mathrm{N}(\mathbf{r,\omega})\otimes
\fo{\mathbf{j}}_\mathrm{N}
(\mathbf{r',\omega'})}=0.
\end{equation}
Using Eqs.~(\ref{2.4b}) and (\ref{2.4b-1}) together with
Eqs.~(\ref{2.2}) and (\ref{2.2a}) and
employing Eqs.~(\ref{2.4c}), (\ref{2.5}), and
(\ref{2.8})--(\ref{2.8b}), we can calculate the
thermal-equilibrium correlation functions of the
electric field and the induction field to obtain
\begin{align}
\label{2.8c}
&\bigl\langle\hat{\mathbf{E}}(\mathbf{r})\otimes
\hat{\mathbf{E}}
(\mathbf{r'})
\bigr\rangle
\nonumber\\
&=\frac{\hbar\mu_{0}}{\pi}
\int_{0}^{\infty}\!\! \D\omega\,\omega^2
\coth\!\left(\frac{\hbar\omega}{2 k_\mathrm{B}T}\right)
\Im \tensor{G}(\mathbf{r,r'},\omega),
\\
\label{2.8d}
&\bigl\langle\hat{\mathbf{B}}(\mathbf{r})\otimes
\hat{\mathbf{B}}
(\mathbf{r'})
\bigr\rangle
\nonumber\\
&=-
\frac{\hbar\mu_{0}}{\pi}
\int_{0}^{\infty}\!\! \D\omega\,
\coth\!\left(\frac{\hbar\omega}{2 k_\mathrm{B}T}\right)
\Nabla\times
\Im \tensor{G}(\mathbf{r,r'},\omega)\times\Lnabla{'}.
\end{align}
Taking the limit $T$ $\!\to$ $\!0$
(i.e., replacing the hyperbolic cotangent with unity)
yields the respective ground-state correlation functions.
Note that the correlation functions (\ref{2.8c}) and (\ref{2.8d})
inherit the reciprocity property according to 
Eq.~(\ref{2.4c}).

Now we calculate the expectation value of the Lorentz force
[which is Hermitean---recall Eqs.~(\ref{eq37-1}) and (\ref{eq37-3})],
\begin{equation}
\label{2.8e}
\mathbf{F} = \int_V \D^3r\,\bigl\langle
\hat{\rho}\hat{\mathbf{E}}
+ \hat{\mathbf{j}}\times\hat{\mathbf{B}}\bigr\rangle, 
\end{equation}
where $\hat{\mathbf{E}}$ and $\hat{\mathbf{B}}$, respectively,
are defined by Eqs.~(\ref{2.4b}) and (\ref{2.4b-1}) 
together with Eqs.~(\ref{2.2}) and (\ref{2.2a}),
and $\hat{\rho}$ and $\hat{\mathbf{j}}$, respectively, are
defined by Eqs.~(\ref{eq34}) and (\ref{eq35}) together
with Eqs.~(\ref{eq36}) and (\ref{eq37}). Following the line suggested
by classical electrodynamics, paying proper attention to
operator symmetrization
as well as regularization, and taking into account that the
time derivative in the (quantum-mechanical version of)
Eq.~(\ref{eq13}) does not contribute to the force, we find
(Appendix~\ref{appA}) that Eqs.~(\ref{eq14}) and (\ref{eq15})
apply, where the (time-independent) stress tensor can be obtained,
in agreement with the classical Eq.~(\ref{Max8}),
from the quantum-mechanical expectation value
\begin{eqnarray}
\label{2.7}
\lefteqn{
\tensor{T}(\mathbf{r,r'})
}
\nonumber\\&&
=\varepsilon_{0}
\EWb{\hat{\mathbf{E}}(\mathbf{r})\otimes\hat{\mathbf{E}}(\mathbf{r'})}
+\mu_{0}^{-1}
\EWb{\hat{\mathbf{B}}(\mathbf{r})\otimes\hat{\mathbf{B}}(\mathbf{r'})}
\nonumber\\&&\hspace{2ex}
-{\textstyle\frac{1}{2}}\,\tensor{1}
\bigl[\varepsilon_{0}\EW{\hat{\mathbf{E}}(\mathbf{r})
\hat{\mathbf{E}}(\mathbf{r'})}
+\mu_{0}^{-1}
\EWb{\hat{\mathbf{B}}(\mathbf{r})\hat{\mathbf{B}}(\mathbf{r'})}
\bigr]
\qquad
\end{eqnarray}
in the limit $\mathbf{r'}\to\mathbf{r}$,
where divergent bulk contributions are to be removed
before taking the limit [recall the remark below Eq.~(\ref{eq15})].
This is always possible if the body under study is   
embedded in a material environment that is
homogeneous at least in the vicinity of the body.
If this is not the case, special care and
additional considerations are necessary, and
it may happen that physically interpretable results
can hardly
be extracted.
Note that in the calculation of the surface integral
in Eq.~(\ref{eq14}) the ``outer'' values of the integrand
should be used if $\partial V$ is the interface between
an \emph{inhomogeneous} body embedded in a
homogeneous environment (see
Appendix~\ref{appA}).

Inserting Eqs.~(\ref{2.8c}) and (\ref{2.8d}) into Eq.~(\ref{2.7})
finally yields the stress tensor as
\begin{equation}
\label{Max17}
\tensor{T}(\mathbf{r,r})=
\lim_{\mathbf{r}'\to\mathbf{r}}\left[
\tensor{\theta}(\mathbf{r,r'})
- {\textstyle\frac{1}{2}} \tensor{1}
{\rm Tr}\,\tensor{\theta}(\mathbf{r,r'})
\right]\!,
\end{equation}
where
\begin{eqnarray}
\label{Max18}
\lefteqn{
\tensor{\theta}(\mathbf{r,r'})
=\frac{\hbar}{\pi}\int_{0}^{\infty} \D\omega\,
\coth\!\left(\frac{\hbar\omega}{2 k_\mathrm{B}T}\right)
}
\nonumber\\&&\hspace{-1ex}\times
\left[\frac{\omega^2}{c^2}\,\Im\tensor{G}(\mathbf{r,r'},\omega)
-\Nabla\times
\Im\tensor{G}(\mathbf{r,r'},\omega)\times\Lnabla{'}
\right]\!.
\qquad
\end{eqnarray}
As expected, the permittivity $\varepsilon(\mathbf{r},\omega)$ 
and the permeability $\mu(\mathbf{r},\omega)$ do not appear 
explicitly in Eq.~(\ref{Max18}), but only via the Green tensor
$\tensor{G}(\mathbf{r,r'},\omega)$.
Having removed divergent bulk contributions, we may take the
imaginary part of the whole integral instead of the integrand
in Eq.~(\ref{Max18}) and rotate the integration contour in the usual
way toward the imaginary frequency axis, on which
the Green tensor is real [recall Eq.~(\ref{2.4d})].
In the zero-temperature limit, the result is simply
\begin{eqnarray}
\label{Max19}
\lefteqn{
\tensor{\theta}(\mathbf{r,r'})
=-\frac{\hbar}{\pi}\int_{0}^{\infty} \D\xi\,
}
\nonumber\\&&\hspace{-1ex}\times
\left[\frac{\xi^2}{c^2}\,\tensor{G}(\mathbf{r,r'},i\xi)
+\Nabla\times
\tensor{G}(\mathbf{r,r'},i\xi)\times\Lnabla{'}
\right]\!.
\qquad
\end{eqnarray}
For nonzero temperatures, a sum over the poles of the hyperbolic
cotangent (corresponding to the Matsubara frequencies)
arises instead. It should be mentioned that the zero-frequency
contribution to the resulting series can be problematic
if the expression in the square bracket in
Eq.~(\ref{Max19}) has a singularity there, which
is the case when permittivities of Drude type
(exhibiting a pole at zero frequency)
are employed. In fact, this unpleasant feature expresses
the conceptual
limitations of a spatially local description of
the material response to the electromagnetic field,
which disregards spatial dispersion. It is known that,
for materials with (almost) freely movable charge carriers,
this can become an issue especially
at low frequencies (large free path lengths).
Extension of the quantization scheme to nonlocally responding
materials would render it possible to include such materials
in the calculation of Casimir forces in a consistent way.

\section{Harmonic-oscillator medium}
\label{sec4}

It is maybe illustrative to make contact with microscopic
approaches to the problem. The simplest and most widely used 
model for describing linearly polarizable media
is quite certainly the harmonic-oscillator model
(inclusion of magnetic properties into the model
is still scarce). To account for dissipation,
the medium oscillators that are
relevant to the linear interaction with the electromagnetic 
field---shortly referred to as medium oscillators---are
also linearly coupled to (infinitely many) heat bath
oscillators (e.g., phonon modes). The
effect of the heat bath can then be adequately
taken into account by including friction terms and associated
noise forces in the equations of motion of the medium oscillators.
On a coarse-grained time scale, the friction terms
are commonly regarded as being local in time
(Markov approximation), 
so that they can be characterized by simple damping constants.
It should be noted that the requirement of limited 
time resolution implies that different, not strictly equivalent
noise forces are acceptable in that regime (for details
of damping theory and oscillator models, see, e.g.,
Refs.~\cite{Lax,GardinerQuantumNoise,WelschQuantumOptics,Ford}). 

In the context of the one-dimensional theory of the Casimir force
on absorbing bodies, the harmonic-oscillator model has been used
to study the interaction of damped medium oscillators with the 
transverse part of the quantized one-dimensional electromagnetic field,
with special emphasis on homogeneous media \cite{Kup}.
Extending the one-dimensional theory
to three dimensions, we begin with
the Heisenberg equations of motion
of the system in the form of
\begin{equation}
\label{Osc2}
\dot{\hat{\mathbf{p}}}(\mathbf{r},t)
=-m\omega_{0}^2
\hat{\mathbf{s}}(\mathbf{r},t)-m \gamma 
\dot{\hat{\mathbf{s}}}(\mathbf{r},t)+e \hat{\mathbf{E}}(\mathbf{r},t)
+\hat{\mathbf{F}}_\mathrm{N}(\mathbf{r},t),
\end{equation}
\begin{equation}
\label{Osc1}
\dot{\hat{\mathbf{s}}}(\mathbf{r},t)
=\hat{\mathbf{p}}(\mathbf{r},t)/m,
\end{equation}
\begin{equation}
\label{Osc3a}
\Nabla\times\hat{\mathbf{B}}(\mathbf{r},t)-
\frac{1}{c^2}\,\dot{\hat{\mathbf{E}}}(\mathbf{r},t)=\mu_{0}
\hat{\mathbf{j}} (\mathbf{r},t),
\end{equation}
\begin{equation}
\label{Osc3b}
\Nabla\times\hat{\mathbf{E}}(\mathbf{r},t)
=-\dot{\hat{\mathbf{B}}}(\mathbf{r},t),
\end{equation}
where $\hat{\mathbf{s}}(\mathbf{r},t)$ and 
$\hat{\mathbf{p}}(\mathbf{r},t)$ are, respectively,
the coordinate field and the momentum field
of the medium oscillators, and
\begin{equation}
\label{eq52}
\hat{\mathbf{j}}(\mathbf{r},t)=
e\eta(\mathbf{r}) \dot{\hat{\mathbf{s}}}(\mathbf{r},t)
\end{equation}
is the (model) current [$\eta(\mathbf{r})$, number density of 
the me\-di\-um oscillators]. Further,
$\hat{\mathbf{F}}_\mathrm{N}(\mathbf{r},t)$ is the Langevin noise for\-ce
acting on the damped harmonic oscillators ($\gamma$,
damping constant). In addition to the equations of motion
Eqs.~(\ref{Osc2})--(\ref{Osc3b})
and the definition (\ref{eq52}),
the commutation relations \cite{Lax}
\begin{equation}
\label{Osc6a}
\left[\hat{\mathbf{s}}(\mathbf{r},t),\hat{\mathbf{p}}(\mathbf{r'},t)
\right]=\frac{i\hbar}{\eta(\mathbf{r})}\,\tensor{\delta}(\mathbf{r-r'})
\end{equation}
and
\begin{equation}
\label{Osc6f}
[\hat{\mathbf{F}}_\mathrm{N}(\mathbf{r},t),
\hat{\mathbf{F}}_\mathrm{N}(\mathbf{r'},t')]
=\frac{2m\gamma}{\eta(\mathbf{r})}\,i\hbar\,\tensor{\delta}
(\mathbf{r-r'})\frac{\partial
\delta(t-t')}
{\partial t}
\end{equation}
together with the standard commutators of the electromagnetic field
are required to specify the model. The commutator (\ref{Osc6f}) 
ensures that Eq.~(\ref{Osc6a}) is preserved in time. Note that 
the proof given in Ref.~\cite{Lax} extends to the
inhomogeneous case $\eta$ $\!=$ $\!\eta(\mathbf{r})$ considered here.
It should be stressed that the number density $\eta(\mathbf{r})$ 
is not allowed to have zeros (nor infinities), otherwise
Eqs.~(\ref{Osc6a}) and (\ref{Osc6f}) were not well-defined. 
Further, note that
\begin{equation}
\label{eq52a}
\hat{\mathbf{E}}(\mathbf{r},t)=
\hat{\mathbf{E}}^{\parallel}(\mathbf{r},t)
+\hat{\mathbf{E}}^{\perp}(\mathbf{r},t)
\end{equation}
is the full electric field consisting of both
longitudinal and transverse parts \footnote{The longitudinal and
   transverse tensorial $\delta$ functions, respectively, are given by
   $\tensor{\delta}^{\parallel}(\mathbf{r})$
   $\!=$ $\![(\Nabla\!\otimes\!\Nabla)/\Delta]\delta(\mathbf{r})$
   and $\tensor{\delta}^{\perp}(\mathbf{r})$ $\!=$ $\!\tensor{\delta}(\mathbf{r})$
   $\!-\tensor{\delta}^{\parallel}(\mathbf{r})$.},
\begin{equation}
\label{eq52d}
\hat{\mathbf{E}}^{\parallel(\perp)}(\mathbf{r},t)
=\int \D^3r'\, \tensor{\delta}^{\parallel(\perp)}(\mathbf{r-r'})
\hat{\mathbf{E}}(\mathbf{r'},t).
\end{equation}
The transverse part may be associated with a vector potential
in the Coulomb gauge and expanded into orthogonal
modes in the usual way. By contrast, the longitudinal part
is not a dynamical electromagnetic field
variable but is (nonlocally) determined
by the oscillator field as
\begin{equation}
\label{eq52b}
\hat{\mathbf{E}}^{\parallel}(\mathbf{r},t)=-e[\eta(\mathbf{r})
\hat{\mathbf{s}}(\mathbf{r},t)]^{\parallel}/ \varepsilon_{0},
\end{equation}
implying the conserved (model) charge density
\begin{equation}
\label{eq52c}
\hat{\rho}(\mathbf{r},t)=-e\Nabla[\eta(\mathbf{r})
\hat{\mathbf{s}}(\mathbf{r},t)],
\end{equation}
which is consistent with Eqs.~(\ref{Osc3a}) and (\ref{eq52}).
Needless to say, Eqs.~(\ref{eq52}) and (\ref{eq52c}) do not
actually represent the sources on a truly microscopic level but rather 
on a mesoscopic one, since the term ``continuously varying field''
applied to matter consisting of well-distinguishable constituents
already indicates some averaging. However, complying with
established terminology, we refer to this mesoscopic description
as being microscopic throughout the paper.
Note that \emph{ab initio} calculations on a truly microscopic level
would lead to time-ordered products in the treatment
of the complicated interaction problem even in linear electrodynamics, 
because of the interaction with the dissipative system.
However, if this interaction is treated in Born and (quasi-)Markov
approximations, then the closed equations derived in this way
no longer contain any time-ordered products.

As the system evolves toward
its dressed ground state as $t$ $\!\to$ $\!\infty$,
the model can be shown (Appendix~\ref{appB}) to lead
to the equal-time electromagnetic field correlation functions
\begin{equation}
\label{Osc30}
\lim_{t\to\infty}
\bigl\langle\hat{\mathbf{E}}(\mathbf{r},t)\otimes
\hat{\mathbf{E}}
(\mathbf{r'},t)
\bigr\rangle
=\frac{\hbar\mu_{0}}{\pi}
\int_{0}^{\infty}\!\! \D\omega\,\omega^2\,
\Im \tensor{G}(\mathbf{r,r'},\omega)
\end{equation}
and
\begin{multline}
\label{Osc31}
\lim_{t\to\infty}
\bigl\langle\hat{\mathbf{B}}(\mathbf{r},t)\otimes
\hat{\mathbf{B}}
(\mathbf{r'},t)
\bigr\rangle
\\
=-
\frac{\hbar\mu_{0}}{\pi}
\int_{0}^{\infty}\!\! \D\omega\,\Nabla\times
\Im \tensor{G}(\mathbf{r,r'},\omega)\times\Lnabla{'}
\end{multline}
if the heat bath that interacts with the medium oscillators
is assumed to have zero temperature.
Here,
$\tensor{G}(\mathbf{r,r'},\omega)$ is the Green tensor that is
the solution to Eq.~(\ref{2.1}), with $\kappa(\mathbf{r},\omega)$ and
$\varepsilon(\mathbf{r},\omega)$, respectively, being the 
model-specific quantities $\kappa(\mathbf{r},\omega)$ $\!\equiv$ $\!1$ and
\begin{equation}
\label{Osc9}
\varepsilon(\mathbf{r},\omega)=1+\frac{e^2\eta(\mathbf{r})}
{\varepsilon_{0}m}\,\frac{1}{\omega_{0}^2-\omega^2-i\gamma\omega}\,.
\end{equation}
Obviously, Eqs.~(\ref{Osc30}) and (\ref{Osc31}), which directly
follow from the microscopic model under consideration, correspond 
exactly to Eqs.~(\ref{2.8c}) and (\ref{2.8d}) in the zero-temperature limit.

The (steady-state) Lorentz force acting on the harmonic-oscillator
matter in some space region $V$ is given by
\begin{equation}
\label{Osc32}
\mathbf{F} = \lim_{t\to\infty}
\int_V \D^3r\,\bigl\langle
\hat{\rho}(\mathbf{r},t)\hat{\mathbf{E}}(\mathbf{r},t)
+ \hat{\mathbf{j}}(\mathbf{r},t)\times\hat{\mathbf{B}}(\mathbf{r},t)
\bigr\rangle,
\end{equation}
with $\hat{\rho}(\mathbf{r},t)$ and $\hat{\mathbf{j}}(\mathbf{r},t)$
from Eqs.~(\ref{eq52c}) and (\ref{eq52}), respectively.
At this stage it not difficult to see that the procedure
outlined in Appendix~\ref{appA} yields $\mathbf{F}$ in 
the form of Eq.~(\ref{eq14}) together with Eq.~(\ref{eq15}),
where the stress tensor has exactly the form of Eq.~(\ref{Max17})
together with Eq.~(\ref{Max19}).
This result shows that the microscopic approach to the Casimir force
fully confirms the macroscopic approach as given in Sec.~\ref{sec3},
where the calculations were based on the quantized macroscopic
electromagnetic field, with the matter phenomenologically
described in terms of Kramers-Kronig consistent
response functions. Thus, the Casimir force acting on a
macroscopic piece of matter may be viewed as ``just'' the
(quantum) Lorentz force on the constituting charges and currents,
which, in a macroscopic description, can be expressed
in terms of the (induced and noise) polarization and magnetization---a
conceptually straightforward and satisfactory point of view.

\section{Casimir force in planar structures}
\label{sec5}

Let us apply the theory to a planar magnetodielectric
structure defined according to
\begin{equation}
\label{L1}
\varepsilon(\mathbf{r},\omega)=
\left\{
\begin{array}{rl}
\varepsilon_{-}(z,\omega),& \quad z<0,\\
\varepsilon_{j}(\omega),& \quad 0<z<d_{j}, \\
\varepsilon_{+}(z,\omega),& \quad z>d_{j},
\end{array}
\right.
\end{equation}
\begin{equation}
\label{L2}
\mu(\mathbf{r},\omega)=
\left\{
\begin{array}{rl}
\mu_{-}(z,\omega),& \quad z<0,\\
\mu_{j}(\omega),& \quad 0<z<d_{j}, \\
\mu_{+}(z,\omega),& \quad z>d_{j}.
\end{array}
\right.
\end{equation}
To determine the Casimir stress in the interspace
\mbox{$0$ $\!<$ $\!z$ $\!<$ $\!d_{j}$},
we need the Green tensor in Eq.~(\ref{Max18}) for both spatial 
arguments within the interspace ($0\!<\!z\!=\!z'\!<\!d_{j}$).
The Green tensor is well known and can be taken, e.g., from
Ref.~\cite{ChewBook}. Since the transverse projection $\mathbf{q}$
of the wave vector is conserved and the polarizations $\sigma$ $\!=$ $\!s,p$ 
decouple, the scattering part of the Green tensor within the interspace
can be expressed in terms of reflection coefficients $r_{j\pm}^{\sigma}$ $\!=$
$\!r_{j\pm}^{\sigma}(\omega,q)$
referring to reflection of waves at the right ($+$) and left ($-$)
wall, respectively, as seen from the interspace.
Explicit (recurrence) expressions for the reflection coefficients 
are available if the walls are multislab magnetodielectrics
like Bragg mirrors \cite{Tomas,ChewBook,Raabe1}. (For
continuous wall profiles, Riccati-type equations have to be solved
\cite{ChewBook}.) In the simplest case of two homogeneous, 
semi-infinite walls, the coefficients $r_{j\pm}^{\sigma}$
reduce to the well-known Fresnel amplitudes. In the
case first treated by Lifshitz \cite{Lifshitz}, the interspace is
empty and the walls are nonmagnetic. 

\subsection{Casimir stress within a nonempty interspace}
\label{sec5a}

For the sake of generality, we first leave the wall structure 
unspecified. By modifying the expression for the scattering part 
of the Green tensor given in Ref.~\cite{Tomas} to
also account for magnetic properties, from 
Eq.~(\ref{Max17}) together with Eq.~(\ref{Max18}) (without the bulk part 
of the Green tensor) it then follows that the relevant stress tensor element
$T_{zz}(\mathbf{r,r})$ in the interspace $0$ $\!<$ $\!z$ $\!<$ $\!d_j$
can be given in the form of
\begin{multline}
\label{L3}
T_{zz}(\mathbf{r,r})=-\frac{\hbar}{8\pi^2}
\int_{0}^{\infty} \D\omega\,
\coth\!\left(\frac{\hbar\omega}{2k_\mathrm{B}T}\right)
\\\times\,
\Re\!\!\int_{0}^{\infty} \D q\,q\,\frac{\mu_{j}(\omega)}
{\beta_{j}(\omega,q)}
\,g_j(z,\omega,q)
\end{multline}
($q$ $\!=$ $\!|\mathbf{q}|$), where the function $g_j(z,\omega,q)$, 
which in general depends on the position $z$ within the interspace, reads
\begin{align}
\label{L4}
&
g_j(z,\omega,q)
\nonumber\\&
= 2\bigl[\beta_{j}^2 (1+n^{-2}_{j})-q^2
(1-n^{-2}_{j})\bigr]
D_{js}^{-1}r_{j+}^{s}r_{j-}^{s}e^{2i\beta_{j}d_{j}}
\nonumber\\
&+
2
\bigl[\beta_{j}^2 (1+n^{-2}_{j})
+q^2 (1-n^{-2}_{j})\bigr]
D_{jp}^{-1}r_{j+}^{p}r_{j-}^{p}e^{2i\beta_{j}d_{j}}
\nonumber\\
&-
(\beta_{j}^2+q^2)(1-n^{-2}_{j})
D_{js}^{-1}\bigl[r_{j-}^{s}
e^{2i\beta_{j} z}+r_{j+}^{s}
e^{2i\beta_{j}(d_{j}-z)}\bigr]
\nonumber\\
&+
(\beta_{j}^2+q^2)(1-n^{-2}_{j})
D_{jp}^{-1}
\bigl[r_{-}^{p}e^{2i\beta_{j} z}+r_{j+}^{p}
e^{2i\beta_{j}(d_{j}-z)}\bigr],
\end{align}
with the definitions
\begin{gather}
\label{L5}
n^2_{j}
=n^2_{j}(\omega)
=\varepsilon_{j}(\omega)\mu_{j}(\omega),\\
\label{L5-1}
\beta_{j}=\beta_{j}(\omega,q)=(\omega^2n_{j}^2/c^2
-q^2)^{1/2},\\
\label{L5-2}
D_{j\sigma}=D_{j\sigma}(\omega,q)=
1-r_{j+}^{\sigma} r_{j-}^{\sigma} e^{2i\beta_{j}d_{j}}.
\end{gather}
Note that the equations $D_{j\sigma}(\omega,q)\!=\!0$
determine, for real $\mathbf{q}$, the frequencies of the
guided waves in the planar structure, which are of major interest in all
``mode summation'' approaches. (In the presence
of material losses, however, these waves have complex 
frequencies and are not ordinary normal modes.) 
For practical reasons, it may be advantageous to transform
the integral over real frequencies in Eq.~(\ref{L3}) into
an integral along the imaginary frequency axis by means of contour
integral techniques [cf. Eqs.~(\ref{Max17}) and (\ref{Max19})].
In particular, in the zero-temperature limit,
Eq.~(\ref{L3}) may be rewritten as
\begin{equation}
\label{L10}
T_{zz}(\mathbf{r,r})=
\frac{\hbar}{8\pi^2}\int_{0}^{\infty}\!\!\D\xi\,
\int_{0}^{\infty} \D q\,q\,\frac{\mu_{j}(i\xi)}
{i\beta_{j}(i\xi,q)}\, g_j(z,i\xi,q).
\end{equation}

 From the derivation it is obvious
that the stress formula (\ref{L3})
[together with Eq.~(\ref{L4})] allows for an arbitrary linear, 
causal interspace medium. 
By contrast, Minkowski's stress tensor [Eq.~(\ref{Max11})]
leads to \cite{TomasCasimir,Raabe1} ($\mu_{j}\equiv 1$)
%
\begin{multline}
\label{L7}
T_{zz}^{(\mathrm{M})}(\mathbf{r,r})
=-\frac{\hbar}{2\pi^2}\int_{0}^{\infty}\!\!\D\omega\,
\coth\!\left(\frac{\hbar\omega}{2k_\mathrm{B}T}\right)\\
\times\,
\Re\!\!\int_{0}^{\infty} \D q\,q \beta_{j}\sum_{\sigma=s,p}
\frac{r_{j+}^{\sigma}r_{j-}^{\sigma}e^{2i\beta_{j}d_{j}}}{D_{j\sigma}}\,.
\end{multline}
 From Eq.~(\ref{L4}) it is easily seen that for an empty
interspace, i.e., $\varepsilon_{j}$ $\!=$ $\!\mu_{j}$ $\!=$ $\!1$,
$g_j(z,\omega,q)$ becomes independent of $z$ and simplifies to 
\begin{equation}
\label{L6}
g_j(z,\omega,q)
\to
g_j(\omega,q) = 
4\beta_{j}^2\sum_{\sigma=s,p}\frac{r_{j+}^{\sigma}
r_{j-}^{\sigma}e^{2i\beta_{j}d_{j}}}{D_{j\sigma}}\,.
\end{equation}
In this case,
and \emph{only} in this case,
Eq.~(\ref{L3}) reduces to Eq.~(\ref{L7}), from which in the case 
of semi-infinite (homogeneous) dielectric walls Lifshitz's well-known 
formula \cite{Lifshitz} can be recovered.
As already mentioned, formulas of the type of Eq.~(\ref{L7}) [which 
need not necessarily be derived within the stress tensor approach 
to the Casimir force] have been claimed to apply
also to the case where the interspace is filled with
dielectric material
\cite{Schwinger,ZhouF071995}, at least
if the material is nonabsorbing
\cite{TomasCasimir}
(see also the textbooks
\cite{AbrikosovStatPhys,GinzburgED,MilonniQuantumVacuum}
and references therein).
Since $T_{zz}^{(\mathrm{M})}(\mathbf{r,r})$ does not depend on the 
position $z$ within the interspace, application of Eq.~(\ref{L7})
implies the very paradoxical result that the force acting on any
slice of material selected within the interspace vanishes identically,
regardless of the presence and arrangement of the remaining material 
(in particular, regardless of the yet unspecified walls).
This unphysical result clearly shows that Eq.~(\ref{L7}) cannot 
be valid if the interspace is not empty, not even if it
may be justified to regard the interspace medium as nonabsorbing.
In contrast, the stress $T_{zz}(\mathbf{r,r})$
obtained from Eq.~(\ref{L3}) [together with Eq.~(\ref{L4})]
is not uniform within an interspace if the interspace is
filled with a medium. Hence it gives rise, in general,
to a nonvanishing force on a slice of interspace material,
and no paradox appears.
 
Let us return to the stress formula (\ref{L3})
[together with Eq.~(\ref{L4})]. It is not difficult to see
that, for a nonempty interspace, the $q$ integral in Eq.~(\ref{L3})
fails to converge at $z$ $\!=$ $\!0$ and \mbox{$z$ $\!=$ $\!d_{j}$},
i.e., on the interfaces where the different materials are
in immediate contact with each other. Mathematically,
the reason for this divergence can be seen in the fact that the 
reflection coefficients obtained under the assumption of
\emph{infinite} lateral extension of the system do not approach zero 
as $q$ tends to infinity. However, large values of $q$ correspond 
to very oblique traveling waves.
In any real planar setup of finite lateral extension,
such high-$q$ waves clearly do not contribute
to the $q$ integral at all; they are not reflected
but walk off instead. Note that a divergence of exactly
the same type already appears also in the standard case of an 
empty interspace in the limit \mbox{$d_{j}$ $\!\to$ $\!0$}.
In order to (approximately) take into account the finite 
lateral extension of an actual planar setup, an appropriately 
chosen cutoff value (depending on the lateral system size)
for the reflection coefficients at high $q$ values
could be introduced, thereby rendering the $q$ integral finite.
Of course, a more satisfactory approach would be to abandon
the translational invariance from the outset, which, however, 
leads to serious mathematical difficulties since
waves with different polarizations and
transverse wave vectors are then no longer decoupled.

Since, according to Eq.~(\ref{eq14}), the Casimir force acting on a body
is given by the integral of the stress tensor over the surface
enclosing the body, the stress tensor on its own is of less importance.
What is really important is the integral force value
over a closed surface. To obtain the force (per unit area) acting on
a (multilayered) plate of infinite lateral extension,
the stress on the two sides of the plate must be taken into
account. As the example in Sec.~\ref{sec5b} shows, it may
then happen that the parts of the stress tensor that
diverge when the plate is approached from the two sides 
cancel each other out. In such a case, the Casimir force
(per unit area)
on a plate remains well defined even if its lateral extension
is assumed to be infinite.

\subsection{Casimir force on a plate in a nonempty cavity}
\label{sec5b}

In order to make contact with recent work on the Casimir
force on bodies embedded in media \cite{TomasCasimir}, let us
calculate the force acting at zero temperature on a homogeneous
plate in a nonempty planar cavity,
%
\begin{figure}[htb]
%
\includegraphics[width=\linewidth]{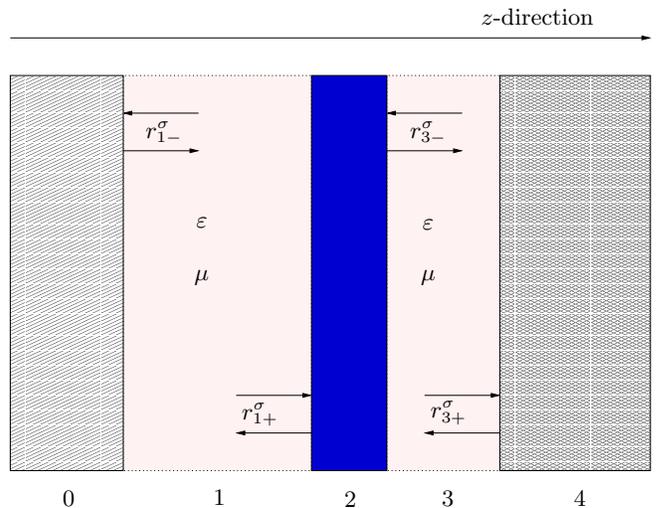}
\caption{\label{SlabInCavityBild}
Homogeneous plate embedded in a nonempty cavity. 
The cavity medium on the right and left sides of the
plate is the same.}
\end{figure}
%
according to the five-region setup as sketched in 
Fig.~\ref{SlabInCavityBild}.
The cavity walls are labeled by $l$ $\!=$ $\!0$ and
$l$ $\!=$ $\!4$, the plate by \mbox{$l$ $\!=$ $\!2$}, and
the cavity regions that are filled with the medium the plate 
is embedded in are labeled
by $l$ $\!=$ $\!1$ and $l$ $\!=$ $\!3$, with
$\varepsilon(\omega)$ $\!\equiv$ $\!\varepsilon_1(\omega)$
$\!=$ $\!\varepsilon_3(\omega)$ and
$\mu(\omega)$ $\!\equiv$ $\!\mu_1(\omega)$
$\!=$ $\!\mu_3(\omega)$.
The total (volume) force per unit transverse area
acting on the plate can be obtained by (vectorial)
addition of the two force contributions
from the two sides of the plate. Application
of Eq.~(\ref{L10}) then yields the total force per unit        
transverse area in the form of 
\begin{multline}
\label{L11a}
F = \frac{\hbar}{8\pi^2}\int_{0}^{\infty}\!\!\D\xi\,
\int_{0}^{\infty} \D q\,q\,
\frac{\mu(i\xi)}{i\beta(i\xi,q)}
\, \bigl[g_3(0,i\xi,q)\\
-g_1(d_1,i\xi,q)\bigr]
\end{multline}
[$\beta(i\xi,q)$ $\!\equiv$ $\!\beta_{1}(i\xi,q)$ $\!=$
$\!\beta_{3}(i\xi,q)$].

For a quantitative comparison with specific results obtained in 
Ref.~\cite{TomasCasimir} on the basis of Minkowski's stress tensor, 
we make the following simplifying assumptions. We assume
that (i) all the reflection coefficients
can be regarded as being almost constant, and (ii)
the reflection coefficients $r_{1+}^{\sigma}$ and
$r_{3-}^{\sigma}$ can be approximated by the (same) single-interface 
(Fresnel) reflection coefficient $r_{1/2}^{\sigma}$.
Physically, these assumptions mean that (i) the distances $d_1$ and $d_3$
between the plate and the cavity walls must not be too small, 
and (ii) the plate must be thick enough. Moreover, the approximation 
scheme implies that the permittivity and the permeability of the medium
the plate is embedded in can be replaced with their static values 
briefly referred to as $\varepsilon$ and $\mu$ in the following, with
$n$ $\!=$ $\!\sqrt{\varepsilon\mu}$ being the static refractive
index. From Eq.~(\ref{L4}) it then follows that the difference of the
functions $g_{3}(0,i\xi,q)$ and $g_{1}(d_1,i\xi,q)$
appearing in Eq.~(\ref{L11a}) can be approximated according to 
\begin{align}
\label{L12}
&g_3(0,i\xi,q) - g_1(d_1,i\xi,q)
\nonumber\\&\quad
\simeq
\sum_{\sigma=s,p}
\Biggl\{
2\biggl(\frac{1}{D_{3\sigma}}-\frac{1}{D_{1\sigma}}\biggr)
\left[\beta^2\left(1+\frac{1}{n^2}\right)
\right.
\nonumber\\&\qquad
\left.+\,\Delta_{\sigma}q^2
\left(1-\frac{1}{n^2}\right)\right]
+\,\Delta_{\sigma}(\beta^2+q^2)\left(1-\frac{1}{n^2}\right)
\nonumber\\&\qquad\ 
\times
\left[\frac{r_{1/2}^{\sigma}+r_{3+}^{\sigma}e^{2i\beta d_{3}}}
{D_{3\sigma}}
-\frac{r_{1/2}^{\sigma}+r_{1-}^{\sigma}e^{2i\beta d_{1}}}
{D_{1\sigma}}\right]
\Biggr\}
\end{align}
($\Delta_{\sigma}$ $\!=$ $\!\delta_{\sigma p}$ $\!-$
$\!\delta_{\sigma s}$), where
\begin{align}
\label{L13}
\frac{ r_{3+}^{\sigma} e^{ 2i\beta d_{3} }}{D_{3\sigma}}
-\frac{r_{1-}^{\sigma} e^{ 2i\beta d_{1} }}{D_{1\sigma}}
&\,=
\frac{1-D_{3\sigma}} {r_{3-}^{\sigma}D_{3\sigma}}
-\frac{1-D_{1\sigma}} {r_{1+}^{\sigma}D_{1\sigma}}
\nonumber\\
&\,\simeq
\frac{1}{r_{1/2}^{\sigma}}
\left(
\frac{1}
{D_{3\sigma}}
-\frac{1}{D_{1\sigma}}
\right).
\end{align}
Substituting Eq.~(\ref{L12}) together with Eq.~(\ref{L13})
into Eq.~(\ref{L11a}), we (approximately) obtain
\begin{multline}
\label{L15}
F = 
\frac{\hbar}{8\pi^2}\int_{0}^{\infty}\!\!\D\xi\,
\int_{0}^{\infty} \D q\,q\,\frac{\mu}
{i\beta}
\sum_{\sigma=s,p}
\left(
\frac{1}
{D_{3\sigma}}
-\frac{1}{D_{1\sigma}}\right)
\\
\biggl\{2\beta^2\left(1+\frac{1}{n^2}\right)
-\Delta_{\sigma}
\frac{\xi^2}{c^2}\left(n^2-1\right)
\biggl(r_{1/2}^{\sigma}+\frac{1}{r_{1/2}^{\sigma}}\biggr)
\\
+2\Delta_{\sigma}q^2 \left(1-\frac{1}{n^2}\right)
\biggr\}.
\end{multline}

 From an inspection of Eq.~(\ref{L15}) it is seen that
there is no divergence; the integrals are well behaved.
It is worth noting that even without application of
the approximation scheme, the integrals in the basic formula 
(\ref{L11a}) do not diverge. The reason is that, for a chosen value of $\xi$, the
coefficients $r_{3-}^{\sigma}(i\xi,q)$ and $r_{1+}^{\sigma}(i\xi,q)$
tend exponentially to the same single-interface Fresnel coefficient
$r_{1/2}^{\sigma}(i\xi,q)$ as $q$ goes to infinity, as may be
seen from relations like
\begin{equation}
\label{L15-1}
r_{1+}^{\sigma}=\frac{r_{1/2}^{\sigma}
+e^{2i\beta_{2}d_{2}}r_{2+}^{\sigma}}{1+r_{1/2}^{\sigma}
e^{2i\beta_{2}d_{2}}r_{2+}^{\sigma}}
\to r_{1/2}^{\sigma}
\ \mathrm{if}\ q\to\infty,
\end{equation}
\begin{equation}
\label{L15-2}
r_{3-}^{\sigma}=\frac{r_{3/2}^{\sigma}
+e^{2i\beta_{2}d_{2}}r_{2-}^{\sigma}}{1+r_{3/2}^{\sigma}
e^{2i\beta_{2}d_{2}}r_{2-}^{\sigma}}
\to r_{3/2}^{\sigma} 
\ \mathrm{if}\ q\to\infty
\end{equation}
together with the relation $r_{3/2}^{\sigma}$ $\!=$
$\!r_{1/2}^{\sigma}$ (valid for arbitrary values of $\xi$ and $q$).
Note that $i\beta_{2}$ $\!\to$ $\!-\infty$ if 
$q$ $\!\to$ $\!\infty$. As a consequence, the divergent contributions to 
the $q$ integral in Eq.~(\ref{L11a}), which would arise
from $g_3(0,i\xi,q)$ and $g_1(d_1,i\xi,q)$ separately,
combine in a convergent fashion. Thus, for the setup under study, 
a $q$ cutoff need not be introduced.

Let us return to Eq.~(\ref{L15}).
If the two walls and the plate are almost perfectly
reflecting, i.e., $r_{1-}^{\sigma}$ $\!\simeq$ $\!r_{3+}^{\sigma}$
$\!\simeq$ $\!\Delta_{\sigma}$,
$r_{1/2}^{\sigma}$ $\!\simeq$ $\!\Delta_{\sigma}$,
then standard evaluation of the integrals leads to
($n$ $\!=$ $\!\sqrt{\varepsilon\mu}$)
\begin{equation}
\label{L16}
F = \frac{\hbar c \pi^2}{240}\,
\sqrt{\frac{\mu}{\varepsilon}}\,
\left(\frac{2}{3}+\frac{1}{3\varepsilon\mu}
\right)
\left(
\frac{1}{d_3^4}
-\frac{1}{d_1^4}\right).
\end{equation}
In particular, if only one wall is present, say the left one,
then Eq.~(\ref{L16}) reduces to ($d_3$ $\!\to$ $\!\infty$,
$d_1$ $\!\equiv$ $\!d$)
\begin{equation}
\label{L16-1}
F = -\frac{\hbar c \pi^2}{240}
\,
\sqrt{\frac{\mu}{\varepsilon}}\,
\left(\frac{2}{3}+\frac{1}{3\varepsilon\mu}
\right)
\frac{1}{d^4}\,,
\end{equation}
which is the generalization of Casimir's well known formula 
\cite{Casimir} for the force between two almost perfectly 
reflecting plates separated by vacuum [$\mu$ $\!=$ $\!\varepsilon$ $\!=$ $\!1$
in Eq.~(\ref{L16-1})] to the case where the interspace between
the plates is filled with
a medium of static permeability $\mu$ and static
permittivity $\varepsilon$.

\begin{figure}[htb]
%
%
%
%
%
%
\includegraphics[width=\linewidth]{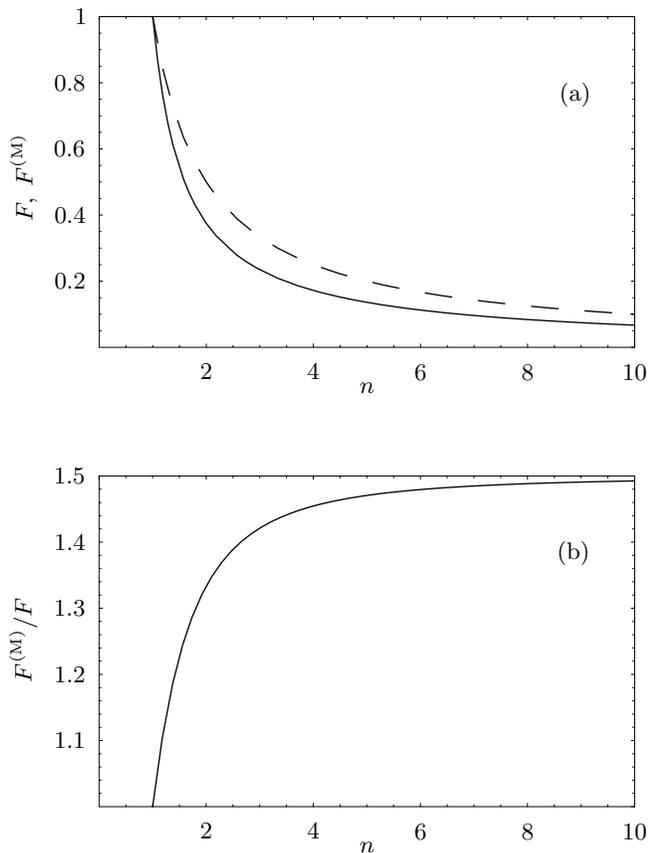}
\caption{%
(a) The Casimir force $F$ given by Eq.~(\ref{L16}) (solid curve) is shown
    as a function of the medium refractive
    index $n$ $\!=$ 
    $\!\sqrt{\varepsilon}$ ($\mu$ $\!=$ $\!1$) for chosen
    distances $d_1$ and $d_3$. For comparison, $F^\mathrm{(M)}$
    given by Eq.~(\ref{L18}) (dashed curve) is shown.    
(b) The ratio $F^\mathrm{(M)}/F$
    is shown as a function of the 
    medium refractive
    index.
}
\label{KoeffizientenBild}
\end{figure}

In order to compare Eq.~(\ref{L16}) with the force
formula obtained on the basis of Minkowski's stress tensor,
we note that the use of Minkowski's stress tensor
for a nonmagnetic medium leads to 
[see Eqs.~(3.6) and (3.7) in Ref.~\cite{TomasCasimir}]
\begin{multline}
\label{L17}
F^\mathrm{(M)} =
-\frac{\hbar}{\pi^2}\int_{0}^{\infty}\D\xi
\int_{0}^{\infty} \D q\,q\,
{i\beta}
\left(
\frac{1}
{e^{-2i\beta d_{3}}-1}
\right.\\\left.
-\,\frac{1}{e^{-2i\beta d_{1}}-1}
\right)
\end{multline}
in place of Eq.~(\ref{L15}) with $\mu$ $\!=$ $\!1$.
For an almost perfectly reflecting plate in a cavity with  
almost perfectly reflecting walls, standard evaluation
of the integrals in Eq.~(\ref{L17}) then yields,
in place of Eq.~(\ref{L16}),
%
\begin{equation}
\label{L18}
F^\mathrm{(M)}
=\frac{\hbar c \pi^2}{240}\,
\frac{1}{\sqrt{\varepsilon}}
\left(
\frac{1}{d_3^{4}}
-\frac{1}{d_1^{4}}\right),
\end{equation}
which in the limit $d_3$ $\!\to$ $\!\infty$ reduces to
($d_1$ $\!\equiv$ $\!d$)
\begin{equation}
\label{L18a}
F^\mathrm{(M)}
=-\frac{\hbar c \pi^2}{240}\,
\frac{1}{\sqrt{\varepsilon}}
\,\frac{1}{d^{4}}\,.
\end{equation}
Note that Eq.~(\ref{L18a}) corresponds
to the result derived in
Ref.~\cite{SchadenM021998}
by means of mode summation methods.
Comparing Eq.~(\ref{L16}) with Eq.~(\ref{L18}) [or Eq.~(\ref{L16-1})
with Eq.~(\ref{L18a})], we see that
\begin{equation}
\label{L18-1}
|F| \leq |F^\mathrm{(M)}|,
\end{equation} 
i.e., the absolute value of the force is ($n$ $\!>$ $\!1$)
always smaller than that predicted from Minkowski's
stress tensor.
Introduction of a (polarizable) medium into the interspace
is obviously associated with some screening of the plate, thereby
reducing the force acting on it. Since the 
internal 
charges and currents
of the interspace medium are not fully taken into account in
a theory that is based on Minkowski's stress tensor or an equivalent
formalism, the screening effect is 
underestimated and consequently the force calculated in this way is
overestimated. Although the assumptions made to derive
the results given above are rather restrictive, the comparison of
Eq.~(\ref{L16}) with Eq.~(\ref{L18}) clearly
shows that the correct inclusion of
the medium into the theory can 
give rise to 
noticeable effects (see Fig.~\ref{KoeffizientenBild}).

A consequence of the approximation scheme employed in this section
is the appearance of the real values of the static permittivity and
the static permeability of the interspace material in Eq.~(\ref{L16}).
However, the basic equation (\ref{L11a}) is of course valid
for arbitrary linear magnetodielectric media with Kramers-Kronig 
consistent permittivities and permeabilities. The influence of
material dispersion and
absorption comes into play when the distances $d_{1}$ and/or $d_{3}$
are decreased. The behavior of the permeability and the
permittivity at nonzero frequencies becomes then important.

\section{Summary and conclusions}
\label{sec6}
On the basis of (i) the quantized macroscopic
electromagnetic field in the presence of
causal linear magnetodielectric media without spatial dispersion
and (ii) the Lorentz force acting on the internal charges and
currents of the medium,
we have derived general expressions 
for the Casimir force acting on magnetodielectric bodies embedded in
a common magnetodielectric medium.
All the matter has been allowed for being dispersing
and absorbing.
Specializing to planar structures,
we have generalized 
Lifshitz-type formulas
(being valid for empty interspaces) to the case where
the interspaces are
filled with a magnetodielectric medium.
In this context, we have analyzed the failure implied by basing
the calculation of the Casimir force on Minkowski's
stress tensor---a method that has been used in the literature but
has never been proven correct.
Interestingly, Lifshitz himself did not address
nonempty interspaces in his seminal article \cite{Lifshitz}.

For comparison reasons,
we have studied in some detail
the Casimir force acting
on a homogeneous plate embedded in a medium
in a planar cavity. Applying
standard approximations such as high reflection, 
we have explicitly
demonstrated that when the plate is embedded in a medium, 
then the force can noticeably differ from
the result obtained on the basis of Minkowski's stress tensor.
By the way, we have given the correct extension 
of Casimir's original formula for the force between
two perfectly reflecting plates to the case where the
interspace between the plates is filled with a medium.

In order to 
make contact with microscopic theories,
we have also described the matter microscopically,
by employing the model of damped harmonic oscillators, which is
widely used for treating dielectric matter.    
Solving the quantum-mechanical equations of motion
of the overall system (with the heat bath assumed in its ground state),
we have calculated the Lorentz force acting on
a chosen matter element. The result obtained in this way
exactly corresponds to the general result obtained
from the macroscopic approach. This clearly shows that
the use of Minkowski's stress tensor to calculate the Casimir force
is wrong in general, even if the matter may be regarded as being
nonabsorbing.

\noindent
\emph{Note added.} 
Instead of Eq.~(\ref{L12}), it may be advantageous 
to use the exact equation
\begin{align}
\label{L12TomaschA}
&g_3(0,i\xi,q) - g_1(d_1,i\xi,q)
\nonumber\\&\quad
=\sum_{\sigma=s,p}
\Biggl\{
2\biggl[\beta^2\left(1+\frac{1}{n^2}\right)
+\,\Delta_{\sigma}q^2
\left(1-\frac{1}{n^2}\right)
\biggr]r^{\sigma}
\nonumber\\&\qquad
+\,\Delta_{\sigma}(\beta^2+q^2)
\left(1-\frac{1}{n^2}\right)(1+r^{\sigma 2}-t^{\sigma 2})
\Biggr\}
\nonumber\\&\qquad\hspace{10ex}\times\;
\frac{r_{3+}^{\sigma}e^{2i\beta d_{3}}
-r_{1-}^{\sigma}e^{2i\beta d_{1}}}
{N^{\sigma}}\,,
\end{align}  
where
\begin{multline}
\label{L12TomaschB}
N^{\sigma}=1-r^{\sigma}(r_{1-}^{\sigma}e^{2i\beta
d_{1}}+r_{3+}^{\sigma}e^{2i\beta d_{3}})
\\
+(r^{\sigma 2}
-t^{\sigma 2})\,r_{1-}^{\sigma}r_{3+}^{\sigma}e^{2i\beta(d_{1}+d_{3})},
\end{multline}
with $r^{\sigma}$ $\!\equiv$ $\!r^{\sigma}_{1/3}$ $\!=$
$\!r^{\sigma}_{3/1}$ and 
$t^{\sigma}$ $\!\equiv$ $\!t^{\sigma}_{1/3}$ $\!=$ $\!t^{\sigma}_{3/1}$
being single-plate reflection and transmission coefficients,
respectively.
We thank Marin-Slo\-bo\-dan Toma\v{s} for this suggestion.

\acknowledgments We thank Ludwig Kn\"{o}ll
and Ho Trung Dung for stimulating and helpful
discussions. C.R. thanks also Mikayel Khanbekyan
for discussions and is grateful for being granted
a Th\"{u}\-rin\-ger Lan\-des\-graduier\-ten\-sti\-pen\-dium.
This work was supported by the
Deutsche Forschungsgemeinschaft.

\begin{appendix}
\section{Proof of Eqs.~(\ref{eq37-1})--(\ref{eq37-4})}
\label{appC}
Using Eqs.~Eqs.~(\ref{2.2}), (\ref{2.2a}), and (\ref{2.1}),
we express $\fo{\rho}(\mathbf{r},\omega)$
and $\fo{j}(\mathbf{r},\omega)$ as defined by
Eqs.~(\ref{eq36}) and (\ref{eq37}), respectively, in terms of
$\fo{j}_\mathrm{N}(\mathbf{r},\omega)$ to obtain
\begin{equation}
\label{C1}
\fo{\rho}(\mathbf{r},\omega) =\frac{i\omega}{c^2}\,
\Nabla\int \D^3r'\,\tensor{G}(\mathbf{r,r'},\omega)
\fo{j}_\mathrm{N}(\mathbf{r'},\omega),
\end{equation}
\begin{equation}
\label{C2}
\fo{j}(\mathbf{r},\omega)
=\left(\Nabla\times\Nabla\times\,-\frac{\omega^2}{c^2}\right)
\int d^3r'\,\tensor{G}(\mathbf{r,r'},\omega)
\fo{j}_\mathrm{N}(\mathbf{r'},\omega).
\end{equation}
By combining Eqs.~(\ref{2.4}), (\ref{2.4-1}), and (\ref{2.4-2})
with the standard bosonic commutation relations for the
fundamental fields $\hat{\mathbf{f}}_\lambda(\mathbf{r},\omega)$ and
$\hat{\mathbf{f}}^\dagger_\lambda(\mathbf{r},\omega)$, it is
not difficult to show that
$\fo{j}_\mathrm{N}(\mathbf{r},\omega)$ and
$\fo{j}_\mathrm{N}{^{\hspace{-1.2ex}\dagger}}\,(\mathbf{r},\omega)$
obey the commutation relation
\begin{align}
\label{C3}
&\bigl[\,\hat{\!\underline{j}}_\mathrm{Nk}(\mathbf{r},\omega),
\,\hat{\!\underline{j}}_\mathrm{Nl}{^{\hspace{-1.8ex}\dagger}}\;
(\mathbf{r'},\omega')\bigr]
=\frac{\hbar}{\mu_{0}\pi}\,\delta(\omega-\omega')
\nonumber\\&
\times \;
\left[
\frac{\omega^2}{c^2}
\sqrt{\Im \varepsilon(\mathbf{r},\omega)}\,
\tensor{\delta}(\mathbf{r\!-\!r'})
\sqrt{\Im \varepsilon(\mathbf{r}',\omega')}
\right.
\nonumber\\&
\left.
+\,\Nabla\!\times\!
\sqrt{\Im \kappa(\mathbf{r},\omega)}\,
\tensor{\delta}(\mathbf{r\!-\!r'})
\sqrt{\Im \kappa(\mathbf{r}',\omega')}\,
\!\times\!\Lnabla{'}
\right]
_{\!kl}
\!
.
\end{align}
 From Eqs.~(\ref{2.2}), (\ref{2.2a}), (\ref{C1}), and (\ref{C2})
together with the commutation relation (\ref{C3}), we derive,
on recalling the Green-tensor relations
(\ref{2.4c}) and (\ref{2.5}),
\begin{align}
\label{C4}
\bigl[\fo{\rho}(\mathbf{r,\omega}),
\fo{E}^{\dagger}(\mathbf{r'},\omega')\bigr]
&= \frac{\hbar}{\pi}\frac{\omega^2}{c^2}\,\delta(\omega-\omega')
\Nabla\,\Im\tensor{G}(\mathbf{r,r'},\omega)
\nonumber\\
&=-\bigl[
\fo{\rho}^{\dagger}(\mathbf{r,\omega}),
\fo{E}(\mathbf{r'},\omega')\bigr],
\end{align}
\begin{align}
\label{C6}
&\bigl[\,\hat{\!\underline{j}}_{k}(\mathbf{r,\omega}),
\hat{\underline{B}}_{l}^{\dagger}(\mathbf{r'},\omega')\bigr]
\nonumber\\&\ 
=-\frac{\hbar}{\pi}\,\delta(\omega\!-\!\omega')
\!\!\left[\!\left(\!\Nabla\!\times\!\Nabla\!\times
-\frac{\omega^2}{c^2}\right)\!
\Im\tensor{G}(\mathbf{r,r'},\omega)
\!\times\!\Lnabla{'}\right]_{\!kl}
\nonumber\\&\ 
=-\bigl[\hat{\underline{j}}_{k}{^{\hspace{-1ex}\dagger}}(\mathbf{r,\omega}),
\hat{\underline{B}}_{l}(\mathbf{r'},\omega')\bigr],
\end{align}
%
\begin{align}
\label{C7-0}
&\bigl[\fo{\rho}(\mathbf{r,\omega}),
\fo{B}^{\dagger}(\mathbf{r'},\omega')\bigr]
\nonumber\\&\
=-\frac{\hbar}{\pi}\,\frac{i\omega}{c^2}\,\delta(\omega-\omega')
\Nabla\,\Im\tensor{G}^{\perp}(\mathbf{r,r'},\omega)\times\Lnabla{'}
\nonumber\\&\
=\bigl[\fo{\rho}^{\dagger}(\mathbf{r,\omega}),
\fo{B}(\mathbf{r'},\omega')\bigr],
\end{align}
\begin{align}
\label{C8-0}
&\bigl[\,\hat{\!\underline{j}}_{k}(\mathbf{r,\omega}),
\hat{\underline{E}}_{l}{^{\hspace{-.9ex}\perp\dagger}}
(\mathbf{r'},\omega')\bigr]
\nonumber\\&
=-\frac{\hbar}{\pi}\,i\omega\,\delta(\omega-\omega')
\left[\!\left(\!\Nabla\!\times\!\Nabla\!\times
-\frac{\omega^2}{c^2}\right)\!
\Im\tensor{G}^{\perp}(\mathbf{r,r'},\omega)\right]_{\!kl}
\nonumber\\
&=\bigl[\,\hat{\!\underline{j}}_{k}^{\dagger}(\mathbf{r,\omega}),
\hat{\underline{E}}_{l}{^{\hspace{-.9ex}\perp}}
(\mathbf{r'},\omega')\bigr],
\end{align}
where
\begin{equation}
\label{C8a}
\tensor{G}^{\perp}(\mathbf{r,r'},\omega)=\int \D^3s\,
\tensor{G}(\mathbf{r,s},\omega)\tensor{\delta}^{\perp}(\mathbf{s-r'}).
\end{equation}
Note that in Eq.~(\ref{C7-0}), $\tensor{G}^{\perp}(\mathbf{r,r'},\omega)$
may be replaced with $\tensor{G}(\mathbf{r,r'},\omega)$,
because of the operation $\times\Lnabla{'}$.

Equations~(\ref{C4}) and (\ref{C6}) obviously imply the
commutation relations
\begin{multline}
\label{C5}
\bigl[\hat{\rho}(\mathbf{r}),
\hat{\mathbf{E}}(\mathbf{r'})\bigr]
=\int_{0}^{\infty} \D\omega
\int_{0}^{\infty}
\D\omega'\,
\biggl\{
\bigl[\fo{\rho}(\mathbf{r,\omega}),
\fo{E}^{\dagger}(\mathbf{r'},\omega')\bigr]
\\
+\bigl[\fo{\rho}^{\dagger}(\mathbf{r,\omega}),
\fo{E}(\mathbf{r'},\omega')\bigr]
\biggr\}
= 0
\end{multline}
and
\begin{multline}
\label{C6b}
\bigl[\,\hat{\!j}_k(\mathbf{r}),
\hat{B}_{l}(\mathbf{r'})\bigr]
=\int_{0}^{\infty} \D\omega
\int_{0}^{\infty}
\D\omega'\,
\biggl\{
\bigl[\,\hat{\!\underline{j}}_{k}(\mathbf{r,\omega}),
\hat{\underline{B}}_{l}^{\dagger}(\mathbf{r'},\omega')\bigr]
\\
+\bigl[\,\hat{\!\underline{j}}_{k}^{\dagger}(\mathbf{r,\omega}),
\hat{\underline{B}}_{l}(\mathbf{r'},\omega')\bigr]
\biggr\}
= 0,
\end{multline}
and hence Eqs.~(\ref{eq37-1}) and (\ref{eq37-3}) are seen to hold.
Note in particular that the commutation relation
$\bigl[\hat{\rho}(\mathbf{r}),
\hat{\mathbf{E}}^{\perp}(\mathbf{r'})\bigr]\!=\!0$ is valid.
 From Eqs.~(\ref{C7-0}) and (\ref{C8-0}), respectively, it follows that
\begin{align}
\label{C7}
&\bigl[\hat{\rho}(\mathbf{r}),
\hat{\mathbf{B}}(\mathbf{r'})\bigr]
=-\frac{2i\hbar}{\pi c^2}\,
\int_{0}^{\infty} \!\!\D\omega\,\omega
\Nabla\,\Im\tensor{G}^{\perp}(\mathbf{r,r'},\omega)\times\Lnabla{'}
\end{align}
and
\begin{align}
\label{C8}
&\bigl[\,\hat{\!j}_{k}(\mathbf{r}),
\hat{E}_{l}^{\perp}(\mathbf{r'})\bigr]
\nonumber\\&\ 
=-\frac{2i\hbar}{\pi}
\int_{0}^{\infty}\!\!\! \D\omega\,\omega
\!\left[\!\left(\Nabla\times\Nabla\times
-\frac{\omega^2}{c^2}\right)
\Im\tensor{G}^{\perp}(\mathbf{r,r'},\omega)\right]_{\!kl}\!.
\end{align}

To further evaluate the integrals in
Eqs.~(\ref{C7}) and (\ref{C8}),
we recall the asymptotic behavior of
$\varepsilon(\mathbf{r},\omega)$
and $\kappa(\mathbf{r},\omega)$ for large $\omega$ in
the upper half-plane, viz.,
\begin{gather}
\label{C14}
\varepsilon(\mathbf{r},\omega) \simeq
1-\frac{\Omega_{\varepsilon}^2(\mathbf{r})}{\omega^2}
%
\,,
\\
\label{C15}
\kappa(\mathbf{r},\omega) 
\simeq
1+\frac{\Omega_{\kappa}^2(\mathbf{r})}{\omega^2}
%
%
\,.
\end{gather}
Substituting Eqs.~(\ref{C14}) and (\ref{C15}) into Eq.~(\ref{2.1}),
we easily see that the Green tensor asymptotically behaves like 
\begin{equation}
\label{C11}
\tensor{G}(\mathbf{r,r'},\omega)
\simeq
-\frac{c^2}{\omega^2}\,\tensor{\delta}(\mathbf{r}-\mathbf{r}')
%
%
\end{equation}
for large $\omega$ in the upper half-plane.
Thus, on recalling Eq.~(\ref{2.4d}) and the holomorphic
behavior of the Green tensor,
we may evaluate the integral in Eq.~(\ref{C7})
to prove Eq.~(\ref{eq37-2}),
\begin{align}
\label{C10}
\bigl[\hat{\rho}(\mathbf{r}),
\hat{\mathbf{B}}(\mathbf{r'})\bigr]
&= -\frac{\hbar}{\pi c^2}\,\mathrm{P}
\int_{-\infty}^{\infty}
\D\omega\,\omega
\Nabla\tensor{G}^{\perp}(\mathbf{r,r'},\omega)\times\Lnabla{'}
\nonumber\\
&= -\frac{\hbar}{\pi c^2}\,
\int_{\mathcal{C}}
\D\omega\,\omega
\Nabla\tensor{G}^{\perp}(\mathbf{r,r'},\omega)\times\Lnabla{'}
\nonumber\\
&= -i\hbar
\Nabla\,\tensor{\delta}^{\perp}(\mathbf{r}-\mathbf{r}')
\times\Lnabla{'} = 0
\end{align}
($\mathrm{P}$ denotes principal value). Here, we have replaced the
principal value integral along the real frequency axis by a
contour ($\mathcal{C}$) integral over an infinitely large
semicircle in the upper half-plane and have used Eq.~(\ref{C11}).
Note that there is no extra pole contribution from $\omega\!=\!0$
\cite{Welsch}. To evaluate Eq.~(\ref{C8}), we take into account 
that, according to Eq.~(\ref{2.1}), the relation
\begin{align}
\label{C16}
&\left(\Nabla\times\Nabla\times
-\frac{\omega^2}{c^2}\right)
\Im\tensor{G}^{\perp}(\mathbf{r},\mathbf{r}',\omega)
\nonumber\\&\qquad
=
\Im\biggl\{
\frac{\omega^2}{c^2}[\varepsilon(\mathbf{r},\omega)-1]
\tensor{G}^{\perp}(\mathbf{r,r'},\omega)
\nonumber\\&\qquad\quad
+\Nabla\times[1-\kappa(\mathbf{r},\omega)]
\Nabla\times
\tensor{G}^{\perp}(\mathbf{r,r'},\omega)
\biggr\}
\end{align}
may be used on the real $\omega$ axis.
Inserting this relation into Eq.~(\ref{C8})
and recalling general
properties of $\varepsilon(\mathbf{r},\omega)$ and
$\kappa(\mathbf{r},\omega)$, we see that the evaluation of
Eq.~(\ref{C8}) can be done in exactly the same way as  
the evaluation of Eq.~(\ref{C7}). Thus, making use of Eqs.~(\ref{C14}),
(\ref{C15}), and (\ref{C11}), we derive
\begin{align}
\label{C17}
\bigl[\,\hat{\!j}_{k}(\mathbf{r}),
\hat{E}_{l}^{\perp}(\mathbf{r'})\bigr]
&= -\frac{\hbar}{\pi}\!
\int_{\mathcal{C}} \D\omega\,\omega\,
\frac{\omega^2}{c^2}[\varepsilon(\mathbf{r},\omega)\!-\!1]
{G}^{\perp}_{kl}(\mathbf{r,r'},\omega)
\nonumber\\
&= i\hbar\,\Omega_{\varepsilon}^2(\mathbf{r})
{\delta}^{\perp}_{kl}(\mathbf{r}-\mathbf{r}'),
\end{align}
which is Eq.~(\ref{eq37-4}).

For a consistency check of the commutation relation (\ref{C17}),
let us consider a set of atoms, with each of them having one
valence electron ($e$, charge; $m$, mass). Let $\mathbf{r}_{A}$
be the (fixed) positions and $\hat{\mathbf{s}}_{A}$
the relative coordinates of the electrons. The microscopic
(electron) current density is then given by
\begin{equation}
\label{C17-1}
\hat{\mathbf{j}}(\mathbf{r})
= e \sum_{A} \dot{\hat{\mathbf{s}}}_{A}
\delta(\mathbf{r}-\mathbf{r}_{A}-\hat{\mathbf{s}}_{A}).
\end{equation}
By assuming minimal coupling and Coulomb gauge, the
canonical momenta of the electrons commute with
the vector potential $\hat{\mathbf{A}}(\mathbf{r})$,
whose conjugate momentum field is
$-\varepsilon_{0}\hat{\mathbf{E}}^{\perp}(\mathbf{r})$.
Hence, we derive
\begin{align}
\label{C18}
&\bigl[\,\hat{\!j}_{k}(\mathbf{r}),
\hat{E}_{l}^{\perp}(\mathbf{r'})\bigr]
\nonumber\\&\quad
= - \frac{e^2}{m} \sum_A
\delta(\mathbf{r}-\mathbf{r}_A-\hat{\mathbf{s}}_A) 
\bigl[\hat{A}_{k}(\mathbf{r}_{A}+\hat{\mathbf{s}}_{A}),
\hat{E}_{l}^{\perp}(\mathbf{r'})\bigr]
\nonumber\\&\quad
= \frac{i\hbar}{\varepsilon_{0}}
\frac{{e^2}}{m}
\sum_A
\delta(\mathbf{r}-\mathbf{r}_A-\hat{\mathbf{s}}_A) 
\delta_{kl}^{\perp}(\mathbf{r}_A+\hat{\mathbf{s}}_A-\mathbf{r'})
\nonumber\\&\quad
= \frac{i\hbar}{\varepsilon_{0}}\,
\frac{{e^2}}{m}
\sum_A
\delta(\mathbf{r}-\mathbf{r}_A-\hat{\mathbf{s}}_A) 
\delta_{kl}^{\perp}(\mathbf{r}-\mathbf{r'}).
\end{align}
In the macroscopic theory, the sum of the $\delta$ functions
in Eq.~(\ref{C18}) is expected to be replaced according to
\begin{equation}
\label{C18-1}
\sum_{A}
\delta(\mathbf{r}-\mathbf{r}_{A}-\hat{\mathbf{s}}_{A})
\mapsto
\sum_{A}
\Delta(\mathbf{r}-\mathbf{r}_{A}-\hat{\mathbf{s}}_{A}),
\end{equation}
where $\Delta(\mathbf{r})$ is a well-behaved function
with unit integral, $\int \D^3r\,\Delta(\mathbf{r})$ $\!=$ $\!1$.
Further, in order to produce reasonable coarse-graining,
$\Delta(\mathbf{r})$ must be sufficiently flat
so that the change of $\Delta(\mathbf{r})$
on atomic length scales can be regarded as being negligibly small. With
the $\hat{\mathbf{s}}_{A}$ acting on well localized electronic
bound states, we may hence write
\begin{equation}
\label{C18-2}
\Delta(\mathbf{r}-\mathbf{r}_A-\hat{\mathbf{s}}_A)
\simeq \Delta(\mathbf{r}-\mathbf{r}_A).
\end{equation} 
Thus,
\begin{equation}
\label{C18-3}
\sum_A
\Delta(\mathbf{r}-\mathbf{r}_A-\hat{\mathbf{s}}_A)
\simeq \sum_A\Delta(\mathbf{r}-\mathbf{r}_A)
= \eta(\mathbf{r}),
\end{equation}
where $\eta(\mathbf{r})$ is
the number density $\eta(\mathbf{r})$ of the atoms, and
the macroscopic version of Eq.~(\ref{C18}) reads
\begin{equation}
\label{C18-4}
\bigl[\,\hat{\!j}_{k}(\mathbf{r}),
\hat{E}_{l}^{\perp}(\mathbf{r'})\bigr]
= \frac{i\hbar}{\varepsilon_0}\,\frac{e^2}{m}\,
\eta(\mathbf{r})
\delta_{kl}^{\perp}(\mathbf{r}-\mathbf{r'}).
\end{equation}
 From a comparison of Eq.~(\ref{C18-4}) with Eq.~(\ref{C17}), 
the relation
\begin{equation}
\label{C18-5}
\Omega^2_{\varepsilon}(\mathbf{r})
= \frac{e^2\eta(\mathbf{r})}{\varepsilon_{0}m}
\end{equation}
is suggested to be valid, which is in full agreement with the
harmonic-oscillator model permittivity given by Eq.~(\ref{Osc9}).

\section{Quantum Lorentz force}
\label{appA}
Using Maxwell's equations (\ref{Max1})--(\ref{Max4})
(promoted to operator equations)
together with the commutation relations
(\ref{eq37-1}) and (\ref{eq37-3}) and relations of the type
\begin{align}
\label{B2}
\hat{\mathbf{E}}(\mathbf{r})
\times & \bigl[\Nabla\times\hat{\mathbf{E}}(\mathbf{r})\bigr]
= -\bigl[\Nabla\times\hat{\mathbf{E}}(\mathbf{r})\bigr]
\times\hat{\mathbf{E}}(\mathbf{r})
\nonumber\\&
=\Nabla\left[{\textstyle\frac{1}{2}}\hat{\mathbf{E}}(\mathbf{r})
\hat{\mathbf{E}}(\mathbf{r})-\hat{\mathbf{E}}(\mathbf{r})
\otimes\hat{\mathbf{E}}(\mathbf{r})\right]
\nonumber\\&\hspace{8ex}
- \bigl[\Nabla\hat{\mathbf{E}}(\mathbf{r})\bigr]
\hat{\mathbf{E}}(\mathbf{r}),
\end{align}
we derive
\begin{align}
\label{B2b}
\hat{\rho}(\mathbf{r})\hat{\mathbf{E}}(\mathbf{r})
&+\hat{\mathbf{j}}(\mathbf{r})\times\hat{\mathbf{B}}(\mathbf{r})
\nonumber\\&
- \bigl[\Nabla \hat{\tensor{T}}(\mathbf{r},\mathbf{r}')\bigl]
_{\mathbf{r'}=\mathbf{r}}
- \bigl[\Nabla' \hat{\tensor{T}}(\mathbf{r},\mathbf{r}')\bigl]
_{\mathbf{r'}=\mathbf{r}}
\nonumber\\[1ex]&
=\varepsilon_{0}\left\{
\begin{array}{r}
\displaystyle\frac{\partial}{\partial t}\,
\bigl[\hat{\mathbf{B}}(\mathbf{r'})
\times
\hat{\mathbf{E}}(\mathbf{r})\bigr]_{\mathbf{r'}=\mathbf{r}}\,,
\\[2ex]
-\displaystyle\frac{\partial}{\partial t}\,
\bigl[\hat{\mathbf{E}}(\mathbf{r})
\times
\hat{\mathbf{B}}(\mathbf{r'})\bigr]_{\mathbf{r'}=\mathbf{r}}\,,
\end{array}
\right.               
\end{align}
where
\begin{align}
\label{B3a}
\hat{\tensor{T}}(\mathbf{r,r'})
&= \varepsilon_{0}\bigl[
\hat{\mathbf{E}}(\mathbf{r})
\otimes
\hat{\mathbf{E}}(\mathbf{r'})
-{\textstyle\frac{1}{2}}\tensor{1}\,
\hat{\mathbf{E}}(\mathbf{r})
\hat{\mathbf{E}}(\mathbf{r'})
\bigr]
\nonumber\\&\quad
+\mu_{0}^{-1}\bigl[
\hat{\mathbf{B}}(\mathbf{r})
\otimes
\hat{\mathbf{B}}(\mathbf{r'})
-{\textstyle\frac{1}{2}}\tensor{1}\,
\hat{\mathbf{B}}(\mathbf{r})
\hat{\mathbf{B}}(\mathbf{r'})
\bigr]
\end{align}
is a reciprocal
operator function of two spatial variables,
\begin{equation}
\label{B3c}
\hat{\tensor{T}}(\mathbf{r,r'})
=\hat{\tensor{T}}{^\mathsf{T}}(\mathbf{r}',\mathbf{r}),
\end{equation}
because of the commutation relations
\begin{equation}
\label{B2a}
\bigl[\hat{\mathbf{E}}(\mathbf{r}),
\hat{\mathbf{E}}(\mathbf{r'})\bigr]=0=
\bigl[\hat{\mathbf{B}}(\mathbf{r}),
\hat{\mathbf{B}}(\mathbf{r'})\bigr].
\end{equation}
Since the left-hand side of Eq.~(\ref{B2b}) is Hermitian,
so is either of the two
alternative right-hand sides,
which means that symmetrization is not necessary.
Thus, Eq.~(\ref{Max9}) is also valid as an operator equation,
and the steady-state equations (\ref{eq14}) and (\ref{eq15}) apply,
with the stress
tensor being defined by Eq.~(\ref{2.7}) in the limit
$\mathbf{r}'$ $\!\to$ $\!\mathbf{r}$. 

To perform the limit, we write the force acting on some
space region $V$ in the form of
\begin{align}
\label{B5}
&
\mathbf{F}
=\lim_{\epsilon \to 0}
\int_{V} \D^3r\int \D^3r'\,
\delta_{\epsilon}(\mathbf{r}-\mathbf{r'})
\nonumber\\&\qquad\times\,
\left\{
\bigl[\Nabla\tensor{T}(\mathbf{r,r'})\bigl]
+ \bigl[\Nabla'\tensor{T}(\mathbf{r,r'})\bigl]
\right\},
\end{align}
where $\delta_{\epsilon}(\mathbf{r}-\mathbf{r'})$
approaches $\delta(\mathbf{r}-\mathbf{r'})$ as $\epsilon$ 
tends to zero. For instance, one could choose
\begin{equation}
\label{B6}
\delta_{\epsilon}(\mathbf{r})=(4\pi\epsilon^2)^{-1}
\delta(|\mathbf{r}|-\epsilon),
\end{equation}
which corresponds to an average over a spherical surface 
of radius $\epsilon$. Let us first consider the case in which the
material properties are homogeneous everywhere, except at the surface
of the volume $V$, where they may change abruptly.
The function $\tensor{T}(\mathbf{r,r'})$ can then be uniquely
decomposed into a bulk part, which is
divergent at \mbox{$\mathbf{r}'$ $\!=$ $\!\mathbf{r}$},
and a scattering part, which is well behaved 
at $\mathbf{r}'$ $\!=$ $\!\mathbf{r}$, and we may write
\begin{align}
\label{B10a}
&\Nabla \tensor{T}^{(\rm scat)}(\mathbf{r,r})
\nonumber\\&\quad
= \bigl[\Nabla\tensor{T}^{(\rm scat)}
(\mathbf{r,r'})\bigl]_{\mathbf{r}'=\mathbf{r}}
+ \bigl[\Nabla'\tensor{T}^{(\rm scat)}
(\mathbf{r,r'})\bigl]_{\mathbf{r}'=\mathbf{r}}.
\end{align}
For the scattering part, the limit \mbox{$\epsilon\to 0$} 
simply restores the $\delta$ function, so Eq.~(\ref{B5}) becomes
\begin{align}
\label{B11}
&\mathbf{F}
= \int_{V} d^3r\, \Nabla\tensor{T}^{(\rm scat)}
(\mathbf{r,r})
\nonumber\\&\qquad
+\lim_{\epsilon \to 0}
\int_{V} \D^3r\int \D^3r'
\delta_{\epsilon}(\mathbf{r}-\mathbf{r'})
\left\{
\bigl[\Nabla\tensor{T}^{(\rm bulk)}(\mathbf{r,r'})\bigl]
\right.
\nonumber\\&\hspace{20ex}
\left.
+ \bigl[\Nabla'\tensor{T}^{(\rm bulk)}(\mathbf{r,r'})\bigl]
\right\}.
\end{align}
The second term on the right-hand side of Eq.~(\ref{B11}), which 
arises from the bulk part, vanishes, as can be seen from the 
following argument \footnote{The argument 
   may be viewed as an application of Curie's principle: 
   ``asymmetrical'' effects are not produced by ``symmetrical'' causes.}.
Since the bulk part
is a function of $\mathbf{r}-\mathbf{r'}$, it follows that
\begin{equation}
\label{B12}
\int_{V} \D^3r \int \D^3r'\,
\delta_{\epsilon}(\mathbf{r}-\mathbf{r'})
\Nabla\tensor{T}^{(\rm bulk)}
(\mathbf{r,r'})=V \mathbf{b}(\epsilon),
\end{equation}
where $\mathbf{b}$ is some vector that depends only on the parameter
$\epsilon$, and in this way selects, somewhat artificially,
a particular direction in space. However, the bulk part
corresponds to a setup where the whole space is filled with 
homogeneous and isotropic material, implying that such a 
preferred direction does not exist,
and we can conclude that $\lim_{\epsilon\to0}\mathbf{b}(\epsilon)$
$\!=$ $\!0$.
To apply the divergence theorem to the first term on the right-hand
side of Eq.~(\ref{B11}) and transform the volume integral
into a surface integral, we note that if the material properties
change discontinuously at the surface $\partial V$ of the chosen
volume $V$ [cf. Eqs.~(\ref{eq36}) and (\ref{eq37})], then
$\tensor{T}^{(\rm scat)}(\mathbf{r,r})$ is also discontinuous there.
In view of the macroscopic description, it is clear that 
the material properties can be regarded as changing
continuously across a sufficiently thin boundary layer.
To include the net change across such a boundary layer,
the ``outer'' values of the integrand should be taken
(indicated by $\partial V_+$),
\begin{eqnarray}
\label{B14}
\mathbf{F}=
\int_{\partial V_+} \D\mathbf{a}\,\tensor{T}^{(\rm scat)}
(\mathbf{r,r}).
\end{eqnarray}
In order to establish the validity of Eq.~(\ref{B14}) 
for the more general case of varying material properties 
inside the chosen space region 
(whose vicinity is again assumed to be homogeneous), 
one has to return to
Eq.~(\ref{B5}) and decompose
$V$ including a thin boundary layer as described above
into sufficiently small, nonintersecting 
cells $V_{i}$. Summing over all cells, one can then show, by using similar arguments as above,
that in the limit of vanishingly small cells,
$V_i$ $\!\to$ $\!0$,
Eq.~(\ref{B14}) is obtained.

\section{Proof of Eqs.~(\ref{Osc30}) and (\ref{Osc31})}
\label{appB}
Combination of Eq.~(\ref{Osc2}) with Eq.~(\ref{Osc1})
yields the second-order differential equation
\begin{equation}
\label{C1a}
m\ddot{\hat{\mathbf{s}}}(\mathbf{r},t)
=-m\omega_{0}^2
\hat{\mathbf{s}}(\mathbf{r},t)-m \gamma 
\dot{\hat{\mathbf{s}}}(\mathbf{r},t)+e \hat{\mathbf{E}}(\mathbf{r},t)
+\hat{\mathbf{F}}_\mathrm{N}(\mathbf{r},t),
\end{equation}
and combination of Eq.~(\ref{Osc3a}) with Eqs.~(\ref{Osc3b})
and (\ref{eq52}) leads to
\begin{equation}
\label{Osc3}
\frac{1}{c^2}\,\frac{\partial^2}{\partial t^2}\hat{\mathbf{E}}(\mathbf{r},t)
+\Nabla\times\Nabla\times\hat{\mathbf{E}}(\mathbf{r},t)
= - e\mu_{0}\eta(\mathbf{r})\ddot{\hat{\mathbf{s}}}(\mathbf{r},t).
\end{equation}
We are interested in the solution to Eqs.~(\ref{C1a}) and 
(\ref{Osc3}) which is reached
in the limit $t$ $\!\to$ $\!\infty$, 
thereby being independent of the initial conditions.
We may represent it in terms of Fourier integrals according to
\begin{equation}
\label{Osc7b}
f(t)=\int_{-\infty}^\infty \D\omega\,e^{-i\omega t}
\underline{f}(\omega)
\Longleftrightarrow
\underline{f}(\omega)=\int_{-\infty}^{\infty}\frac{\D t}{2\pi}
\,e^{i\omega t} f(t).
\end{equation}
Note that the $\omega$ integrals should be treated as principal
value integrals (with respect to $\omega$ $\!=$ $\!0$) if necessary.
 From Eqs.~(\ref{C1a}) and (\ref{Osc3}) it follows that
the Fourier transforms $\fo{s}(\mathbf{r},\omega)$ and
$\fo{E}(\mathbf{r},\omega)$ of $\hat{\mathbf{s}}(\mathbf{r},t)$
and $\hat{\mathbf{E}}(\mathbf{r},t)$, respectively, are determined by
\begin{equation}
\label{Osc7}
\fo{s}(\mathbf{r},\omega)
=
[m\omega_{0}^2\!-\!m\omega^2\!-\!im\gamma\omega]^{-1}
[\fo{F}_\mathrm{N}(\mathbf{r},\omega)+e\fo{E}(\mathbf{r},\omega)]
\end{equation}
and
\begin{equation}
\label{Osc8}
\Nabla\times\Nabla\times\fo{E}(\mathbf{r},\omega)
-\frac{\omega^2}{c^2}\,
\fo{E}(\mathbf{r},\omega)
=e\mu_{0}\eta(\mathbf{r})\omega^2{\fo{\mathbf{s}}}(\mathbf{r},\omega).
\end{equation}
Substituting $\fo{s}(\mathbf{r},\omega)$ from Eq.~(\ref{Osc7}) into
Eq.~(\ref{Osc8})
and rearranging,
we obtain
\begin{equation}
\label{Osc14}
\Nabla\!\times\!\Nabla\!\times\!\fo{E}
(\mathbf{r},\omega)-\frac{\omega^2}{c^2}\varepsilon(\mathbf{r},
\omega)\fo{E}(\mathbf{r},\omega)
=i\mu_{0}\omega
\fo{J}_\mathrm{N}(\mathbf{r},\omega),
\end{equation}
where $\varepsilon(\mathbf{r},\omega)$, which is given by 
Eq.~(\ref{Osc9}), defines the permittivity of the 
harmonic-oscillator medium, and
\begin{equation}
\label{Osc10}
\fo{J}_\mathrm{N}(\mathbf{r},\omega)
= -\frac{i\omega \varepsilon_{0}}{e}
\left[\varepsilon(\mathbf{r},\omega)-1\right]
\fo{F}_\mathrm{N}(\mathbf{r},\omega)
\end{equation}
is the current density associated with the Langevin force. The
unique inversion of Eq.~(\ref{Osc14}) is
\begin{equation}
\label{Osc20a}
\fo{E}(\mathbf{r},\omega)
=i\mu_{0}\omega\int \D^3r'\,
\tensor{G}(\mathbf{r,r'},\omega)
\fo{J}_\mathrm{N}(\mathbf{r'},\omega),
\end{equation}
where $\tensor{G}(\mathbf{r,r'},\omega)$ is the Green tensor that, for
$\kappa(\mathbf{r,\omega})$ $\!=$ $\!1$ 
and $\varepsilon(\mathbf{r,\omega})$ from Eq.~(\ref{Osc9}),
solves Eq.~(\ref{2.1}) together with the boundary condition at infinity.

To prove Eq.~(\ref{Osc30}), we write
\begin{align}
\label{Osc24a}
&\lim_{t\to\infty}
\EW{\hat{\mathbf{E}}(\mathbf{r},t)\otimes \hat{\mathbf{E}}
(\mathbf{r'},t)}
\nonumber\\&\quad 
=\lim_{t\to\infty}\int_{-\infty}^{\infty}\!\!\D\omega
\!\int_{-\infty}^{\infty} \!\!\D\omega'
e^{-i(\omega+\omega')t}
\EW{\fo{E}(\mathbf{r},\omega)\otimes
\fo{E}(\mathbf{r'},\omega')},
\end{align}
where, according to Eq.~(\ref{Osc20a}) 
[together with Eq.~(\ref{2.4c})],
\begin{eqnarray}
\label{Osc24}
\EW{\fo{E}(\mathbf{r},\omega)\otimes
\fo{E}(\mathbf{r'},\omega')}
= -\mu_{0}^{2}\omega\omega'\!\int\! \D^3s\!\int\!\D^3s'
\tensor{G}(\mathbf{r,s},\omega)
\nonumber\\
\times \,
\EW{\fo{J}_\mathrm{N}(\mathbf{s},\omega)\otimes 
\fo{J}_\mathrm{N}(\mathbf{s'},\omega')}
\tensor{G}(\mathbf{s',r'},\omega').
\quad\quad
\end{eqnarray}
If the heat bath is in the vacuum state, then
\begin{multline}
\label{Osc6d}
\EW{\fo{F}_\mathrm{N}(\mathbf{r},\omega)\otimes
\fo{F}_\mathrm{N}(\mathbf{r'},\omega')}
=\frac{m\gamma
\hbar}{\pi\eta(\mathbf{r})}\,\tensor{\delta}(\mathbf{r-r'})
\\\times\,
\int_{0}^{\infty}\D\omega''\,
\omega''\delta(\omega''-\omega)\delta(\omega''+\omega')
\end{multline}
holds \cite{Lax}, and we find, on recalling Eq.~(\ref{Osc10}),
\begin{align}
\label{Osc26}
&\EW{\fo{J}_\mathrm{N}(\mathbf{s},\omega)\!\otimes\! 
\fo{J}_\mathrm{N}(\mathbf{s'},\omega')}
=-\frac{\omega\omega'\varepsilon_{0}^2}{e^2}\,
[\varepsilon(\mathbf{s},\omega)\!-\!1]
[\varepsilon(\mathbf{s'},\omega')\!-\!1]
\nonumber\\&
\ \times
\frac{m\gamma
\hbar}{\pi\eta(\mathbf{s})}\,\tensor{\delta}(\mathbf{s\!-\!s'})
\!\int_{0}^{\infty}\!\!\D\omega''\,
\omega''\delta(\omega''\!-\!\omega)\delta(\omega''\!+\!\omega').
\end{align} 
Combining Eqs.~(\ref{Osc24a}), (\ref{Osc24}), and (\ref{Osc26}),
we derive
\begin{align}
\label{Osc27}
&\lim_{t\to\infty}
\EW{\hat{\mathbf{E}}(\mathbf{r},t)\otimes \hat{\mathbf{E}}
(\mathbf{r'},t)}=\frac{m\gamma\hbar}{\pi e^2 c^4}
\int_{0}^{\infty}\D\omega\,\omega^5
\nonumber\\&
\ \times\!
\int \!\D^3s\, \tensor{G}(\mathbf{r,s},\omega)\frac{
[\varepsilon(\mathbf{s},\omega)\!-\!1]
[\varepsilon(\mathbf{s},-\omega)\!-\!1]
}{\eta(\mathbf{s})}\,
\tensor{G}(\mathbf{s,r'},-\omega).
\end{align}
 From Eq.~(\ref{Osc9}) it follows that the relation
\begin{equation}
\label{Osc28}
\frac{[\varepsilon(\mathbf{s},\omega)-1]
[\varepsilon(\mathbf{s},-\omega)-1]
}{\eta(\mathbf{s})}=\frac{e^2}{\varepsilon_{0}m\gamma}
\,\frac{\Im \varepsilon(\mathbf{s},\omega)}{\omega}
\end{equation}
is valid for real $\omega$.
Hence, we may rewrite Eq.~(\ref{Osc27}) as
\begin{multline}
\label{Osc29}
\lim_{t\to\infty}
\EW{\hat{\mathbf{E}}(\mathbf{r},t)\otimes \hat{\mathbf{E}}
(\mathbf{r'},t)}=\frac{\hbar\mu_{0}}{\pi}
\int_{0}^{\infty}\D\omega\,\frac{\omega^4} {c^2}
\\\times \,
\int \D^3s\, \tensor{G}(\mathbf{r,s},\omega)
\,\Im \varepsilon(\mathbf{s},\omega)
\tensor{G}(\mathbf{s,r'},-\omega),
\end{multline}
which by means of Eqs.~(\ref{2.4d}) and (\ref{2.5})
eventually leads to Eq.~(\ref{Osc30}).

To calculate
\begin{align}
\label{Osc24a-1}
&\lim_{t\to\infty}
\EW{\hat{\mathbf{B}}(\mathbf{r},t)\otimes \hat{\mathbf{B}}
(\mathbf{r'},t)}
\nonumber\\&\quad 
=\lim_{t\to\infty}\int_{-\infty}^{\infty}\!\!\D\omega
\!\int_{-\infty}^{\infty} \!\!\D\omega'
e^{-i(\omega+\omega')t}
\EW{\fo{B}(\mathbf{r},\omega)\otimes
\fo{B}(\mathbf{r'},\omega')},
\end{align}
we express $\EW{\fo{B}(\mathbf{r},\omega)\otimes
\fo{B}(\mathbf{r'},\omega')}$ in terms of $\EW{\fo{E}(\mathbf{r},\omega)
\otimes\fo{E}(\mathbf{r'},\omega')}$, by using Eq.~(\ref{Osc3b}) 
in the Fourier domain,
\begin{equation}
\label{Osc30b}
\fo{B}(\mathbf{r},\omega)=(i\omega)^{-1}
\Nabla\times\fo{E}(\mathbf{r},\omega).
\end{equation}
By means of Eqs.~(\ref{Osc24}), (\ref{Osc26}), and (\ref{Osc28})
[together with Eqs.~(\ref{2.4d}) and (\ref{2.5})] it is now
not difficult to prove Eq.~(\ref{Osc31}). Note that there
are no problems at $\omega$ $\!=$ $\!0$.
\end{appendix}
%

\bibliography{StressBib}
\end{document}